\documentclass[12pt]{article}
\usepackage{amsfonts}
\usepackage{amssymb}
\usepackage{amsmath} 
\usepackage[margin=1in]{geometry}
\usepackage{graphicx,xcolor}
\usepackage[english]{babel} 
\usepackage{blindtext}
\usepackage[compress]{cite}
\usepackage{booktabs}
\usepackage{multicol}
\usepackage{tabularx}
\usepackage{diagbox}

\usepackage{algorithm}
\usepackage[noend]{algpseudocode}
\makeatletter
\def\BState{\State\hskip-\ALG@thistlm}
\makeatother
\newcommand{\N}{\mathbb{N}}
\newcommand{\R}{\mathbb{R}}

\newtheorem{definition}{Definition}

\usepackage{wrapfig}
\usepackage{tcolorbox}

\newcommand*\samethanks[1][\value{footnote}]{\footnotemark[#1]}
\title{\LARGE The Confluence of Networks, Games and Learning\footnote{Prepared for IEEE control system magazine, as part of the special issue ``Distributed Nash Equilibrium Seeking over Networks''. } \\ \Large A game-theoretic framework for multi-agent decision making over networks}
\author{Tao Li\footnote{Corresponding author} \thanks{Department of Electrical and Computer Engineering, New York University, NY, USA; Email: \tt{\{tl2636, gp1363, qz494\}@nyu.edu}.}, Guanze Peng\samethanks, Quanyan Zhu\samethanks, Tamer Ba\c{s}ar\thanks{Department of Electrical and Computer Engineering \&  Coordinated Science Laboratory, University of Illinois at Urbana-Champaign, IL, USA; Email: \tt{\{basar1\}@illinois.edu}.}}
\date{}

\newif\ifPDF \ifx\pdfoutput\undefined\PDFfalse \else\ifnum\pdfoutput > 0\PDFtrue \else\PDFfalse \fi \fi
\ifPDF 
\usepackage[pdftex, plainpages = false, colorlinks=true, linkcolor=black, citecolor = green!50!blue, urlcolor = blue, filecolor=black, pagebackref=false, hypertexnames=false,  pdfpagelabels,draft ]{hyperref}
\fi
\usepackage{bbm}
\usepackage[compress]{cite}
\DeclareMathOperator*{\argmax}{arg\,max}
\DeclareMathOperator*{\argmin}{arg\,min}
\usepackage{pifont}
\usepackage{xcolor}

\usepackage{cleveref}
\makeatletter
\def\endthebibliography{%
  \def\@noitemerr{\@latex@warning{Empty `thebibliography' environment}}%
  \endlist
}
\makeatother

\begin{document}
\maketitle
\begin{abstract}
    Recent years have witnessed significant advances in technologies and services in modern network applications, including smart grid management, wireless communication, cybersecurity as well as multi-agent autonomous systems. Considering the heterogeneous nature of networked entities, emerging network applications call for game-theoretic models and learning-based approaches in order to create distributed network intelligence that responds to uncertainties and disruptions in a dynamic or an adversarial environment.  This paper articulates the confluence of networks, games and learning, which establishes a theoretical underpinning for understanding multi-agent decision-making over networks.  We provide an selective overview of game-theoretic learning algorithms within the framework of stochastic approximation theory, and associated applications in some representative contexts of modern network systems, such as the next generation wireless communication networks, the smart grid and distributed machine learning. In addition to existing research works on game-theoretic learning over networks, we highlight several new angles and research endeavors on learning in games that are related to recent developments in artificial intelligence.  Some of the new angles extrapolate from our own research interests. The overall objective of the paper is to provide the reader a clear picture of the strengths and challenges of adopting game-theoretic learning methods within the context of network systems, and further to identify fruitful future research directions on both theoretical and applied studies. 
\end{abstract}

\section{Introduction}\label{sec:intro}
Multi-agent decision making over networks has recently attracted an exponentially growing number of researchers from the systems and control community. The area has gained increasing momentum in various fields including engineering, social sciences, economics, urban science, and artificial intelligence, as it serves as a prevalent framework for studying large and complex systems, and has been widely applied in tackling many problems arising in these fields, such as social networks analysis \cite{jackson2010social}, smart grid management\cite{maharjan2013dependable,zhu2012differential}, traffic control \cite{groot2014toward}, wireless and communication networks \cite{han2012game,zhu2012interference,han_niyato_saad_basar_2019}, cybersecurity \cite{manshaei2013game,zhu2015game}, as well as multi-agent autonomous systems \cite{Stone2000}. 

Due to the proliferation of advanced technologies and services in modern network applications, solving the decision-making problems in multi-agent networks calls for novel models and approaches that can capture the following characteristics of emerging network systems and the design of autonomous controls:
\begin{enumerate}
    \item the heterogeneous nature of the underlying network, where multiple entities, represented by the set of nodes, aim to pursue their own goals with independent decision-making capabilities;
    \item the need for distributed or decentralized operation of the system, when the underlying network is of a complex topological structure and is too large to be managed in a centralized approach; 
    \item the need for creating network intelligence that is responsive to changes in the network and the environment, as the system oftentimes operates in a dynamic or an adversarial environment.  
\end{enumerate}
Game theory provides a natural set of tools and frameworks addressing these challenges, and bridging networks to decision making. It entails development of mathematical models that both qualitatively and quantitively depict how the interactions of self-interested agents with different information and rationalities can attain a global objective or lead to emerging behaviors at a system level. Moreover, with the underlying network, game-theoretic models capture the impact of the topology on the process of distributed decision making, where agents plan their moves independently according to their goals and local information available to them, such as their observations of their neighbors.   

In addition to game-theoretic models over networks, learning theory is indispensable when designing decentralized management mechanisms for network systems, in order to equip networks with distributed intelligence. Through the combination of game-theoretic models and associated learning schemes, such network intelligence allows heterogeneous agents to interact strategically with each other and learn to respond to uncertainties, anomalies, and disruptions, leading to desired collective behavior patterns over the network or an optimal system-level performance. The key feature of such network intelligence is that even though each agent's own decision-making process is influenced by the others' decisions, the agents reach an equilibrium state, that is, a Nash equilibrium as we elucidate later, in an online and decentralized manner. To equip networks with distributed intelligence, networked agents should adapt themselves to the dynamic environment with limited and local observations over a large network that may be unknown to them. Computationally, decentralized learning scales efficiently to large and complex networks, and requires no global information regarding the entire network, which is more practical compared with centralized control laws. 

\begin{figure}[h]
    \centering
    \includegraphics[width=0.6\textwidth]{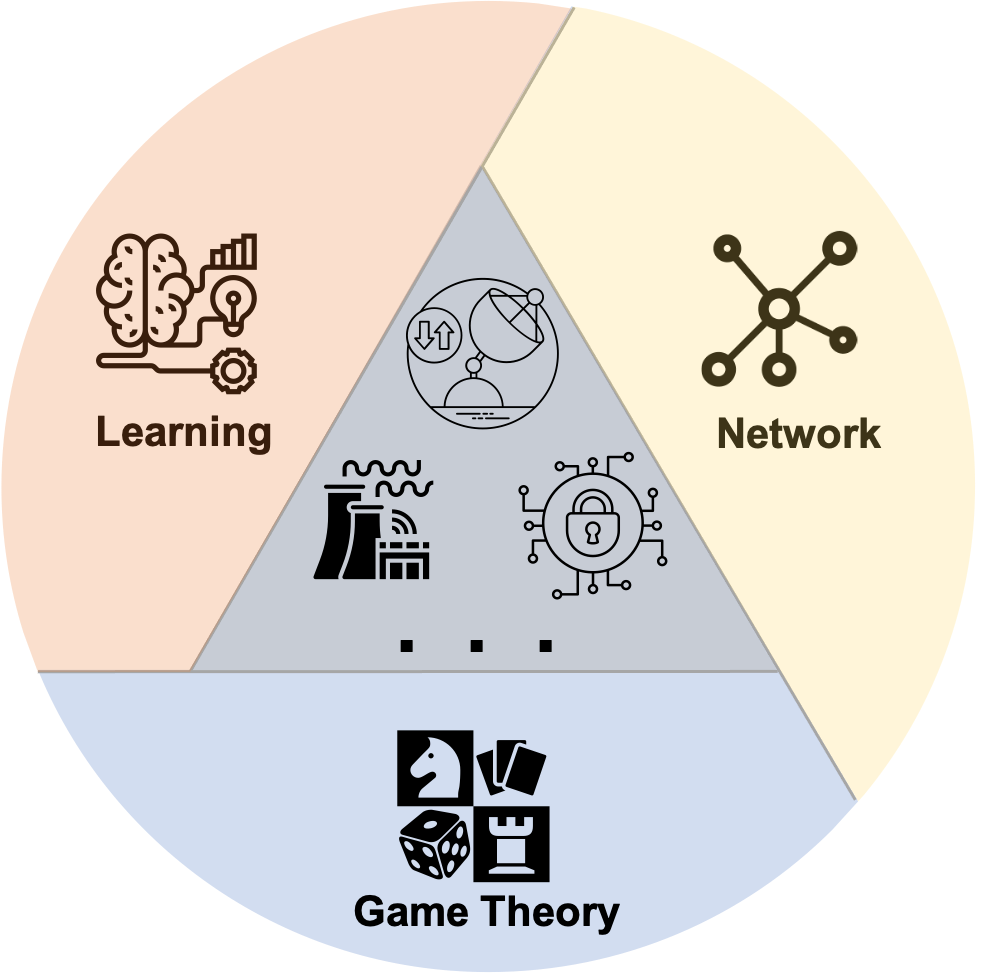}
    \caption{The confluence of networks, games and learning. The combination of game-theoretic modelling and learning theories leads to resilient and agile network controls for various networked systems.}
    \label{fig:game_net}
\end{figure}    

This paper articulates the confluence of networks, games and learning, which establishes a theoretical underpinning for understanding multi-agent decision-making over networks.  We aim to provide a systematic treatment of game-theoretic learning methods and their applications in network problems,  which meet the three requirements specified above.  As shown in \Cref{fig:game_net}, emerging network applications call for novel approaches, and thanks to the decentralized nature, game-theoretic models as well as associated learning methods provide an elegant approach for tackling network problems arising from various fields.  Specifically, our objectives are threefold:
\begin{enumerate}
    \item to provide a high-level introduction to game-theoretic models that apply to multi-agent decision making problems;
    \item to present the key analytical tool based on stochastic approximation and Lyapunov theory for studying learning processes in games, and pinpoint some extensively studied learning dynamics;
    \item to introduce various multi-agent systems and network applications that can be addressed through game-theoretic learning.
\end{enumerate}

We aim to provide the reader a clear picture of the strengths and challenges of adopting novel game-theoretic learning methods within the context of network systems. Besides the highlighted contents, we also provide the reader with references for further reading.
In this paper, complete-information games are the basis of the subject, for which we give a brief introduction to both static and dynamics games. More comprehensive treatments on this topic as well as other game models, such as incomplete information games, can be found in \cite{basar,fudenbergbook,maschler_solan_zamir_2013}.
As most of the network topologies can be characterized by the structure of the utility function of the game \cite{jackson2010social,jackson15network_games}, we do not articulate the influence of network topologies on the game itself. Instead, we focus on its influence on the learning process in games, where players' information feedback depends on the network structures, and we present representative network applications to showcase this influence. We refer the reader to \cite{jackson2010social,jackson15network_games} for further reading on games over various networks.

We structure our discussions as follows. In \Cref{sec:noncoop}, we introduce non-cooperative games and associated solution concepts, including Nash equilibrium and its variants, which capture the strategic interactions of self-interested players. Then, in \Cref{sec:learning}, we move to the main focus of this paper: learning dynamics in games that converge to  Nash equilibrium. Within the stochastic approximation framework, a unified description of various dynamics is provided, and the analytical properties can be studied by ordinary differential equation (ODE) methods. In \Cref{sec:learn_net}, we discuss applications of these learning algorithms in networks, leading to distributed and learning-based controls for network systems. Finally, \Cref{sec:concl}  concludes the paper. For the reader's convenience, we summarize the notations that are frequently used  in \Cref{tab:notation}.

\begin{table}[t]
\centering
\begin{tabular}{ll}
	\toprule
	Symbol & Meaning\\
	\midrule
	$\mathcal{N}$ & The set of players\\
	$i,j\in \mathcal{N}$ & Subscript index denoting players\\
	$\mathcal{N}(i)$ & The set of neighbors of player $i$ \\
	$\mathcal{A}_i$& The set of actions available to player $i$\\
	$\Delta(\mathcal{A}_i)$ & The set of Borel probability measures\\
	&(The probability simplex in $\R^{\mathcal{A}_i}$ for finite action set $\mathcal{A}_i$)\\ 
	$s\in \mathcal{S}$ &  State variable\\
	$u_i:\prod_{j\in N} A_j\rightarrow\R$ & Player $i$'s utility function\\
	$a_i\in \mathcal{A}_i$& Action of player $i$ \\
	$a_{-i}\in \prod_{j\in N, j\neq i} A_j$ & Joint actions of players other than $i$\\
	$\mathbf{a}\in \prod_{i\in \mathcal{N}}\mathcal{A}_i$ & Joint actions of all players\\
	$\pi_i\in \Delta(\mathcal{A}_i)$& Strategy of player $i$  \\
	$\pi_{-i}\in \prod_{j\in N, j\neq i}\Delta(\mathcal{A}_j)$& Joint strategy of players other then $i$\\
	$\mathbf{u}_i(\pi_{-i})$ or $\mathbf{u}_i\in \R^{|\mathcal{A}_i|}$ & Player $i$'s utility vector in finite games\\
	$D_i(\mathbf{a})$ & Individual payoff gradient of player $i$\\
	$D(\mathbf{a})$ & The concatenation of $\{D_i(\mathbf{a})\}_{i\in \mathcal{N}}$\\
	$I_i^k$ & Feedback of player $i$ at time $k$\\
	%$V^\pi_i(s)$ & State value function of player $i$ under strategy profile $\pi$\\
	%$Q^\pi_i(s,a)$ & State-action value function or $Q$ function of player $i$ under strategy profile $\pi$ \\ 
	%$\hat{Q}_i(s,a)$ & Estimator of player $i$' s $Q$ function\\ 
	$U_i^k\in \R$ & The payoff feedback received by player $i$ at time $k$\\
	$\hat{\mathbf{u}}_i^k\in \R^{|\mathcal{A}_i|}$ &  Estimated utility vector at time k\\
	$\hat{\mathbf{U}}^k_i\in \R^{|\mathcal{A}_i|}$ & Estimator of $\mathbf{u}_i(\pi_{-i}^k)$ at time $k$ \\
	  $BR_i$ & Best response mapping for player $i$\\
	  $QR^\epsilon$ & Regularized best response or quantal response \\

	  \bottomrule
	  \end{tabular}

\caption{Table of Notations}
\label{tab:notation}	
\end{table}
 
\section{Noncooperative Game Theory}\label{sec:noncoop}
Game theory constitutes a mathematical framework with two main branches: noncooperative game theory and cooperative game theory. Noncooperative game theory focuses on the strategic decision-making process of independent entities or players that aim to optimize their distinct objective functions, without any external enforcement of cooperative behaviors. The term noncooperative does not necessarily mean that players are not engaged in cooperative behaviors. As a matter of fact, induced cooperative or coordinated behaviors do arise in noncooperative circumstances, within the context of Nash equilibrium, a solution concept of  noncooperative games. However, such coordination is self-enforcing and arises from decentralized decision-making processes of self-interested players, and will be further discussed in \Cref{sec:learn_net}, where we introduce game-theoretic methods for distributed machine learning.

As briefly discussed above, noncooperative game theory naturally characterizes the decision-making process of heterogeneous entities acting independently over networks, which is the main focus of this paper. In the following, we introduce various game models and related solution concepts, including Nash equilibrium and its variants. Generally speaking, a game involves the following elements: decision makers (players); choices available to each player (actions); knowledge that a player acquires for making decisions (information) and each player's preference ordering among its actions, affected by also others' actions (utilities or cost). Below we provide a short list of these concepts that will be further discussed and explained in this section.
\begin{enumerate}
	\item \textit{Players} are participants in a game, where they compete for their own good. A player can be an individual or encapsulation of a set of individuals.
	\item \textit{Actions} of a player, in the terminology of control theory, are the implementations of the player's control.
	\item \textit{Information} in games refers to the structure regarding the knowledge players acquire about the game and its history when they decide on their moves. The information structure can vary considerably. For some games, the information is \textit{static} and does not change during the play. While for other games, new information will be revealed after players' moves, as the “state” of the game, a concept to be elucidated later, is determined by players' actions during the play. In the latter case, the information is \textit{dynamic}. We shall address both types of games in this paper.   
	\item \textit{A strategy} is a mapping that associates a player’s move with the information available to him at the time when he decides which move to choose. 
	\item \textit{A utility or payoff} is oftentimes a real-valued function capturing a player's preference ordering among possible outcomes of the game. Using the terminology in control theory, this can also be viewed as a cost function for the player's controls. 
\end{enumerate}

The above list refers to elements of games in relatively imprecise common language terms, and more formal  definitions are presented below. To facilitate this discussion, we categorize noncooperative games into two main classes: static and dynamic games, based on the nature of the information structure. 
\subsection{Static Games}
Static games are one-shot, where players make decisions simultaneously based on the prior information on the games, such as sets of  players' actions, and their payoffs. In such games, each player's knowledge about the game is static and does not evolve during the play. Mathematically speaking, a static noncooperative game is defined as follows.
\begin{definition}[Static Games]\label{def:static}
	A static game is defined by a triple $G:=\left\langle \mathcal{N}, (\mathcal{A}_i)_{i\in \mathcal{N}}, (u_i)_{i\in \mathcal{N}}\right\rangle$, where
	\begin{enumerate}
		\item $\mathcal{N}=\{1,2,\dots,N\}$ is a finite set of players;
		\item $\mathcal{A}_i$ with some specified topology denotes the set of actions available to the player $i\in \mathcal{N}$; 
		\item $u_i:\prod_{j\in \mathcal{N}} \mathcal{A}_j\rightarrow\R$ defines player $i$'s utility, and  $u_i(a_i, a_{-i})$ gives the payoff of player $i$ when taking action $a_i$, given other players' actions $a_{-i}:= (a_j)_{j\in N, j\neq i}$. 
	\end{enumerate}
\end{definition}

In static games, each player develops its strategy, a probability distribution over his action set, with the objective of maximizing the expected value of its own utility. If players have finite action sets, then such a static game is called a finite game. In this case, a strategy is a finite-dimensional vector in the probability simplex over the action set, that is, $\pi_i\in\Delta(\mathcal{A}_i):=\{\pi\in \mathbb{R}^{|\mathcal{A}_i|}| \pi(a)\geq 0, \forall a \in\mathcal{A}_i, \sum_{a\in \mathcal{A}_i}\pi(a)=1 \}$.
If $\pi_i$ is a unit vector $e_a, a\in \mathcal{A}_i$ with the $a$-th entry being 1 and 0 for others, then it is referred to a pure strategy, selecting action $a$ with probability 1; otherwise, it is a mixed strategy, choosing actions randomly  under the selected probability distribution. Similarly, for infinite action sets, the strategy is defined as a Borel probability measure over the action set, with Dirac measure being the pure strategy. By a possible abuse of notation, we denote the set of Borel probability measures over $\mathcal{A}_i$ by $\Delta(\mathcal{A}_i)$. Unless specified otherwise, static games considered in this paper are all assumed to be finite, where the player set, and the action sets are all finite.

As a special case of games with infinite actions, the mixed extension of finite games is introduced in the sequel. Consider a two-player finite game $G=\left\langle \mathcal{N}, (\mathcal{A}_i)_{i\in \mathcal{N}}, (u_i)_{i\in \mathcal{N}}\right\rangle$, where $\mathcal{N}=\{1,2\}$, and the action sets are finite $|\mathcal{A}_i|<\infty, i\in \mathcal{N}$. Given the mixed strategies of players, $\pi_i\in \Delta(\mathcal{A}_i)$, the expected utility of player $i$ is $\mathbb{E}_{a_1\sim \pi_1,a_2\sim\pi_2}[u_i(a_1,a_2)]$. With a slight abuse of notation, we denote this expected utility by $u_i(\pi_1,\pi_2):=\mathbb{E}_{a_1\sim \pi_1,a_2\sim\pi_2}[u_i(a_1,a_2)]$. Then, studying the players' strategic interactions is equivalent to considering the following infinite game $G^\infty=\left\langle \mathcal{N}, (\Delta(\mathcal{A}_i))_{i\in \mathcal{N}}, (u_i)_{i\in \mathcal{N}}\right\rangle$, where $u_i$ denotes the expected utility. In $G^\infty$, an action is a vector from the corresponding probability simplex, a convex and compact set with a continuum of elements.  Similar to the notations used in the definition, for the mixed extension $G^\infty$, we denote the joint action of players other than $i$ by $\pi_{-i}:=(\pi_j)_{j\in \mathcal{N}, j\neq i	}$. Furthermore, we let $\mathbf{u}_i(\pi_{-i})\in \R^{|\mathcal{A}_i|}$ be the utility vector of player $i$, given other players' strategy profiles, $\pi_{-i}$, whose $a$-th entry is defined as $\mathbf{u}_{i}(\pi_{-i})(a):=u_{i}(e_a,\pi_{-i})$. Due to the definition of expectation,  $u_i(\pi_i,\pi_{-i})$ can be expressed as an inner product $\langle \pi_i, \mathbf{u}_i(\pi_{-i})\rangle$, which will be frequently used later when discussing learning algorithms in finite games. This mixed extension allows us to give a geometric characterization to Nash equilibria of finite games, based on variational inequalities, as discussed in \Cref{sec:solu}. Meanwhile, this inner product expression connects learning theory in finite games with online linear optimization \cite{shai_online}, where the generic player's decision variable is $\pi_i$ and the loss function specified by $\langle \cdot, \mathbf{u}_i(\pi_{-i})\rangle$ is linear in $\pi_i$.

Even though widely applied in modeling behaviors of self-interested players, the static game model is far from being sufficient to cover multi-agent decision making problems arising in different fields. For instance, when playing poker games, new information will be revealed during the game play, such as cards played at each round, based on which players can adjust their moves. There are many games where players' information about the game changes over time during the play, which cannot be suitably described by static games. Therefore, we must resort to another model for capturing the underlying dynamics.

\subsection{Dynamic Games}\label{sec:dynamic_game}
To explicitly represent the dynamic nature of the decision-making process, we adopt system theory terminology and use the state of the game to describe its evolution over a period of time, which could be finite or infinite. Roughly speaking, the current state specifies the current situation of the dynamic game, including the set of players who are about to take actions, actions available to them and their utilities at this time. The fundamental difference between static games and dynamic games is that for the latter the game changes over time as players implement their sequences of actions during the play. Hence, players' knowledge regarding the game also evolves, as players can fully or partially observe the current state. 

In the following, a subclass of Markov games is introduced as an example of dynamic games, which is a very popular game model for studies on multi-agent sequential decision making under uncertainties, such as multi-agent reinforcement learning \cite{littman94markov_game}. 
\begin{definition}[Markov Games]\label{def:markov}
	An $N$-person discrete-time infinite horizon discounted Markov game  consists of
	\begin{enumerate}
		\item a player set $\mathcal{N}=\{1, 2,\ldots, N\}$;
		\item a discrete time set $\N_{+}:= \{1, 2, \cdots\}$, with actions by players taken at each $k\in \N_{+}$;
		\item a set $\mathcal{A}_i$ with some specified topology, defined for each $i\in \mathcal{N}$, corresponding to the set of actions or controls available to player $i$;
		\item a set $\mathcal{S}$ with some specified topology, denoting the state space of the game, where $s^k\in \mathcal{S}, k\in \N_{+}$ represent the state of the game at time $k$;
		\item a transition kernel $T:\mathcal{S}\times\prod_{i\in \mathcal{N}}\mathcal{A}_i\rightarrow \Delta(\mathcal{S})$, according to which the next state is sampled, that is, $s^{k+1}\sim T(s^k,\mathbf{a}^k)$, where $\mathbf{a}^k=(a^k_1,\ldots,a^k_N)$ is the $N$-tuple of  actions  at time $k\in\N_{+}$, and $s^1\in {\cal S}$ has a given distribution;
		\item an instantaneous payoff:  $u_i:  \mathcal{S}\times\prod_{i}\mathcal{A}_i\rightarrow\R$, defined for each $i\in \mathcal{N}$ and $k\in \N_{+}$,  determining the payoff $u_i(s^k,\mathbf{a}^k)$ received by player $i$ at time $k$  ;
		\item a discounting factor $\gamma$.  Given $\{s^1,\ldots,s^k,\dots;\mathbf{a}^1,\ldots,\mathbf{a}^k,\ldots\}$,  the discounted cumulative payoffs for player $i$ is $\sum_{k=1}^\infty \gamma^k u_i(s^k,\mathbf{a}^k)$. 
	\end{enumerate}
\end{definition} 
The above definition only characterizes one special case of dynamic games. Based on this definition, we can derive many other game models. For example, we can make state transitions independent of players' actions as well as the current state, yielding a special case of stochastic games, which will be further discussed in another paper in this special issue \cite{lei_csm}.  
We can also consider continuous-time dynamic games where the transition is described by a differential equation, leading to a differential game model. For an extensive coverage of  dynamic game models, we refer the reader to \cite{basar}.  
 
 With the full observation of states, we can consider the stationary strategy $\pi_i: \mathcal{S}\rightarrow\Delta(\mathcal{A}_i)$, by which players plan their moves only based on the current state $s\in \mathcal{S}$. In this case, we say the state variable $s$ characterizes players' knowledge of the game, since the actions, utilities and next possible states are all determined by the current state.  For dynamic games under partial observation and/or non-Markovian transition,  we refer the reader to \cite{basar}, since these topics are beyond the scope of this paper.   
 
\subsection{Solution Concepts}\label{sec:solu}
The solution or outcome of any given game is more or less a matter of understanding game rules and relations between players. However, besides these concrete  matters, there exist general principles, which dictate players' behaviors and apply to all games. Here, we argue that these principles revolve around the notion of rationality, based on which we introduce the solution concept of Nash equilibrium and some of its variants. Mathematically speaking, a solution to an $N$-person game is a collection of all players' strategies, that has attractive properties expressed in terms of payoffs received by the players. In addition, players can admit different strategies depending on how the game is defined and, in particular, on the information that players acquire. We start with static games, where the information structure is relatively simple. 

Compared with single-agent optimization problems, the analysis of games is more involved, as each player's utility is determined not only by its own decision but also by others' moves. Hence, when a player takes an action, it must take into account possible moves of the other players, which leads to the notion of best response. To introduce ``best response", for clarity, but without any loss of conceptual generality, let us focus on games with two players.  For player 1, given the other player's strategy $\pi_2$, the optimal choice is
\begin{align}\label{eq:br_nash}
	\pi_1&\in BR_1(\pi_2):=\argmax_{\pi\in\Delta(\mathcal{A}_1)}\{\langle\pi,\mathbf{u}_1(\pi_{2})\rangle\},
\end{align}
which is referred to as a best response of player $1$ to player $2$'s strategy $\pi_2$, and $BR_1(\cdot)$ is called the best response set of player 1. Similarly, given  player 1's strategy $\pi_1$, a best response of player 2 is $\pi_2\in BR_2(\pi_1):=\argmax_{\pi\in\Delta(\mathcal{A}_2)}\{\langle\pi,\mathbf{u}_2(\pi_{1})\rangle\}$.
Therefore, we can define a point-to-set mapping $BR:\Delta(\mathcal{A}_1)\times\Delta(\mathcal{A}_2)\rightarrow 2^{\Delta(\mathcal{A}_1)\times\Delta(\mathcal{A}_2)}$, which is the concatenation of $BR_1$ and $BR_2$. Given a joint strategy profile $\pi=(\pi_1,\pi_2)$, $BR(\pi)$ is defined as 
\begin{align}\label{eq:br_mapping}
	BR(\pi):=\{({\pi'}_1,{\pi'}_2)|{\pi'}_1\in BR_1(\pi_2),{\pi'}_2\in BR_2(\pi_1) \}.
\end{align} 
If we can find $\pi^*=(\pi^*_{1},\pi^*_{2})$, a fixed point of this best-response mapping, that is,  $\pi^*\in BR(\pi^*)$, then when both players adopt the corresponding strategy in this profile, they could do no better by unilaterally deviating from current strategy. In other words, this fixed point corresponds to an equilibrium outcome of the game, which further leads to the definition of Nash equilibrium,  which we introduce below for the general $N$-player game. 
\begin{definition}[Nash Equilibrium]
	For a static game $\left\langle \mathcal{N}, (\mathcal{A}_i)_{i\in \mathcal{N}}, (u_i)_{i\in \mathcal{N}}\right\rangle$, Nash equilibrium is a strategy profile $\pi^*=(\pi_i^*,\pi_{-i}^*)$ with the property that for all $i\in \mathcal{N}$,
	\begin{align}\label{eq:ne}
		u_i(\pi_{i}^*, \pi_{-i}^*)\geq u_i(\pi_{i},\pi_{-i}^*),
	\end{align}
	where $\pi_{i}$ is an arbitrary strategy of player $i$ and $\pi^*_{-i}=(\pi^*_j)_{j\in \mathcal{N}, j\neq i	}$ denotes the joint strategy profile of the other players. If the inequality holds strictly for all $\pi_i\neq \pi^*_i$, then  it is referred to as a strict Nash equilibrium.
\end{definition}

\begin{figure}
    \centering
    \includegraphics[width=0.5\textwidth]{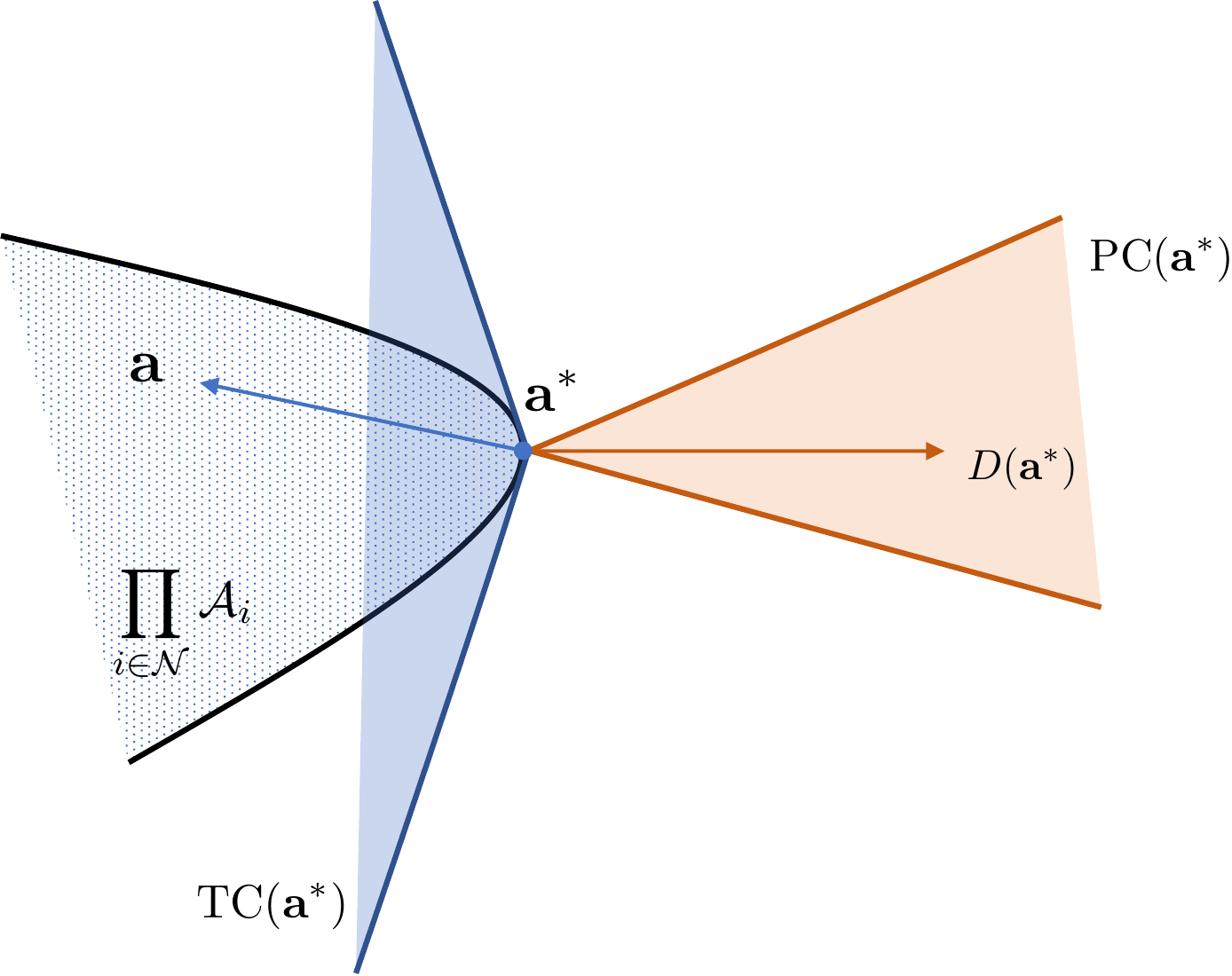}
    \caption{Variational characterization of a Nash equilibrium $\mathbf{a}^*$ in concave games. TC($\mathbf{a}^*$) and PC($\mathbf{a}^*$) denote, respectively, the tangent and the polar cone of $\prod_{i\in \mathcal{N}}\mathcal{A}_i-\mathbf{a}^*$. According to the variational inequality \eqref{eq:vi_ne}, $\mathbf{a}^*$ is a Nash equilibrium if and only if $D(\mathbf{a}^*)$ lies in the polar cone. }
    \label{fig:vi_ne}
\end{figure}

Note that the preceding definition naturally carries over to games with infinite action sets, and we refer the reader to \cite[Chapter 4]{basar} for more details. Furthermore, for infinite games, if we impose some topological structures on the action sets and regularity conditions on the utility functions, we can come up with a geometric interpretation of Nash equilibrium derived from the inequality in \eqref{eq:ne}. Toward that end, we consider a (static) game with compact and convex action sets $(\mathcal{A}_i)_{i\in \mathcal{N}}$ and smooth concave utilities:
\begin{align*}
   u_i(a_i,a_{-i}) \text{ is concave in } a_{i} \text{ for all }a_{-i}\in \prod_{j\in \mathcal{N}, j\neq i}\mathcal{A}_j, i\in \mathcal{N}. 
\end{align*}
In such a game, the number of actions to each player is a continuum, and the utility function is continuous;  such games is referred to as continuous-kernel games or continuous games. In this case, a pure strategy Nash equilibrium $\mathbf{a}^*=(a_i^*,a_{-i}^*)\in \prod_{i\in \mathcal{N}}\mathcal{A}_i$ is defined by the following inequality, 
\begin{align}\label{eq:concave_ineq}
    u_i(a_i^*,a_{-i}^*)\geq u_i(a_i,a_{-i}^*),\quad \text{for all } a_i\in \mathcal{A}_i \text{ and all } i\in \mathcal{N}.
\end{align}
Further assuming that $u_i(a_i, a_{-i})$ is continuously differentiable in $a_i\in {\cal A}_i$, for all $a_{-i}$,  by the first order condition, Nash equilibrium in \eqref{eq:concave_ineq} can be characterized by  
\begin{align*}
    \langle D_i(\mathbf{a}^*), a_i-a_i^*\rangle \leq 0,\quad \text{ for all }a_i\in \mathcal{A}_i, i\in \mathcal{N}, 
\end{align*}
where $D_i(\mathbf{a}):=\nabla_{a_i}u_i(a_i,a_{-i})$ denotes the individual payoff gradient of  player $i$, and $\nabla_{a_i}u_i(a_i,a_{-i})$ denotes differentiation with respect to the variable $a_i$. By rewriting the inequality above in a more compact form, we obtain the following variational characterization  of Nash equilibrium  
\begin{align}\label{eq:vi_ne}
    \langle D(\mathbf{a}^*), \mathbf{a}-\mathbf{a}^*\rangle\leq 0, \quad\text{for all } \mathbf{a}\in \prod_{i\in \mathcal{N}}\mathcal{A}_i,
\end{align}
where $D(\mathbf{a})$ is the concatenation of $\{D_i(\mathbf{a})\}_{i\in \mathcal{N}}$, that is, $D(\mathbf{a})=(D_1(\mathbf{a}),\ldots,D_N(\mathbf{a}))$. Geometrically speaking, \eqref{eq:vi_ne} states that for concave games, $\mathbf{a}^*$ is a Nash equilibrium if and only if $D(\mathbf{a}^*)$ lies within the polar cone of the set $\prod_{i\in \mathcal{N}}\mathcal{A}_i-\mathbf{a}^*:=\{ \mathbf{a}-\mathbf{a}^*|\mathbf{a}\in \prod_{i\in \mathcal{N}}\mathcal{A}_i\}$, as shown in Fig~\ref{fig:vi_ne}.  

%If the utility functions are not concave, the variational inequality \eqref{eq:vi_ne} only determines the game's critical point, i.e, an action profile $\mathbf{a}\in \prod_{i\in \mathcal{N}}\mathcal{A}_i$ for which an infinitesimal unilateral deviation cannot increase the payoff  \textcolor{red}{say more}.  
In addition to concave games, such variational inequality characterization has been studied in much broader contexts, such as monotone games \cite{rosen65concave}, which bridges the gap between the theory of monotone operators and Nash equilibrium seeking. For a detailed discussion, we refer the reader to another paper in this special issue \cite{pavel_csm}. The variational inequality \eqref{eq:vi_ne} is referred to as the Stampacchia-type inequality in the literature \cite{pappallardo02nash_vi}, and a similar variational inequality of this type can also be derived in the context of the mixed extension. As a special case of continuous games, the mixed extension of finite games also satisfies the regularity conditions: the action spaces are probability simplex regions, which are compact and convex, and the utility function, due to its linearity with respect to any player's mixed strategy, is naturally smooth and concave. Therefore, the mixed strategy Nash equilibrium can be characterized by variational inequality as well. Thanks to the inner product expression of the utility in the mixed extension, the individual payoff gradient is simply $\mathbf{u}_i(\pi_{-i})$, and we denote the concatenation of $\{\mathbf{u}_i\}_{i\in \mathcal{N}}$ by $\mathbf{u}(\pi):=[\mathbf{u}_1(\pi_{-1}(t)),\mathbf{u}_2(\pi_{-2}(t)),\ldots, \mathbf{u}_N(\pi_{-N}(t))]$, which we also refer to as the joint utility vector under the strategy profile $\pi$. In the same spirit of \eqref{eq:vi_ne}, a strategy profile $\pi^*$ is Nash equilibrium of the underlying finite game if and only if the following Stampacchia-type inequality holds 
\begin{align}\label{eq:finite_nash_vi}
      \langle \mathbf{u}(\pi^*), \pi-\pi^*\rangle \leq 0, \quad \text{for all } \pi\in \prod_{i\in \mathcal{N}} \Delta(\mathcal{A}_i). \tag{SVI}
\end{align}
As we will later see in \Cref{sec:nash_lyap}, this variational characterization of Nash equilibrium bridges the equilibrium concept of games and the equilibrium concept of dynamical systems induced by learning algorithms.

In the same spirit of \eqref{eq:ne},  Nash equilibrium in dynamic games can also be defined accordingly. For Markov games, given players' stationary strategy profile $\pi$, the cumulative expected utility of player $i$, starting from the initial state $s^1=s$, is 
\begin{align}\label{eq:v_func}
	V_i^\pi(s):=\mathbb{E}_{s^{k+1}\sim T, \mathbf{a}^k\sim \pi}[\sum_{k=1}^\infty \gamma^k u_i(s^k,\mathbf{a}^k)|s^1=s],
\end{align}
which is referred to as state-value function in Markov decision process \cite{puterman_mdp}. If we view $V^\pi_i$ as a function of the strategy profile, following \eqref{eq:ne}, we can define Nash equilibrium for the Markov game, where the inequality holds for every state. In other words, regardless of previous play, as long as players follow $\pi^*$ from the current state $s$, they achieve the best outcome for the rest of the game, and no player has any incentive to deviate from the strategy dictated by $\pi^*$. Hence, this kind of Nash equilibrium is referred to as subgame perfect Nash equilibrium (SPNE), which is widely used in the study of dynamic games \cite{selten1975reexamination,BASAR19899}.
%is of great interest to both researchers and practitioners when working on social-economic problems \cite{Wei17MPE}, multi-agent autonomous systems \cite{kaiqing_overview} and many other related problems.    

The Nash equilibrium serves as a building block for noncooperative games. One of its major advantages is that it characterizes a stable state of a noncooperative game, in which no rational player has the incentive to move unilaterally. This stability idea will be further discussed when we focus on learning in games, which relates stability theory of differential equations to the convergence of learning algorithms in Nash equilibrium seeking.

\section{Learning in Games}\label{sec:learning}
Learning in games refers to a long-run non-equilibrium process of learning, adaptation, and/or imitation that leads to some equilibrium \cite{fuden_learning}. Different from pure equilibrium analysis based on the definition, learning in games accounts for how players behave adaptively during repeated game play under uncertainties and partial observations. Computationally speaking, computing NE   based on equilibrium analysis is challenging due to the computational complexity \cite{nash_complexity}, and this hardly accounts for the decision-making process in practice, where players have limited computation power and information. Hence, learning models are needed to describe how less than fully rational players behave in order to reach the equilibrium. Equilibrium seeking or computation motivates learning in games \cite{BASAR19899}. 

If we view the learning process as a dynamical system, then the learning model can predict how each player adjusts its behavior in response to other players over time to search for strategies that will lead to higher payoffs. From this perspective, a Nash equilibrium can also be interpreted as the steady state of the learning process, which serves as a prediction of the limiting behavior of the dynamical system induced by the learning model. This viewpoint has been widely adopted in the study of population biology and evolutionary game theory, as we shall see more clearly when we discuss later reinforcement learning and replicator dynamics \cite{taylor_replicator}.
 
In this section, various learning dynamics are presented in the context of infinitely repeated games for Nash equilibrium seeking. We consider a number of players repeatedly playing the game $\left\langle \mathcal{N}, (\mathcal{A}_i)_{i\in \mathcal{N}}, (u_i)_{i\in \mathcal{N}}\right\rangle$ infinitely many times. At time $k$, players determine their moves based on their observations up to time $k-1$. Then, they receive feedback from the environment, which provides information on the past actions. For example, in finite games, based on the information available to it, player $i$ constructs a mixed strategy $\pi^k_i\in \Delta(\mathcal{A}_i)$, from which it samples an action $a^k_i$ and implements it. Then it will receive a payoff feedback related to  $u_i(a^k_{i},a^k_{-i})$, which evaluates the performance of $a_i^k$ and helps the player shape its strategy for future plays. In such a repeated game, the amount of information that players acquire in repeated plays directly determines how players plan their moves at each round and further influences the resulting learning dynamics. Besides being of theoretical importance, the information feedback in the learning process, such as players' observations of their opponents' moves, is also of vital importance in designing learning-based methods for solving network problems. As we shall see more clearly in \Cref{sec:learn_net}, in many network applications, networked agents only observe their surroundings, without any access to the global information regarding the whole network. Therefore, due to its significance in learning processes, we first present existing feedback structures that are of wide use in learning, before moving to the details of learning algorithms.
 
\subsection{Feedback Structures in Learning}
The feedback structure for a player in a repeated game includes its observations regarding the game and the repeated plays, which is a subset of every player's histories of plays and payoffs. To make our discussion more concrete, we introduce the following notation.  Let $I_i^k$ be the feedback of player $i$ up to time $k$. Denote the payoff received by player $i$ at the $k$-th round by $u_i^k:=u_i(a_i^k,a_{-i}^k)$, and the sequence of payoffs received up to time $k$ by $u_i^{1:k}:=\{u_i^1,\ldots, u_i^{k}\}$.

The simplest feedback structure is called the \textit{perfect global feedback}, where  $$I_i^k=\{\{u_j^{1:k}\}_{j\in \mathcal{N}},\{a_j^{1:k}\}_{j\in \mathcal{N}}\},$$ indicating the completeness of the feedback from both the temporal and the spatial sense. Furthermore, we can also consider the noisy feedback of payoffs, $U_i^k$, defined as
$$U_i^k=u_i(a_i^k,a_{-i}^k)+\xi_i^k,$$
where $\xi_i^k$ is a zero-mean martingale noise process with finite second moment, that is $\mathbb{E}[\xi_i^k|\mathcal{F}^{k-1}]=0,  \mathbb{E}[(\xi_i^k)^2|\mathcal{F}^{k-1}]$ is less than a constant, and the expectation is taken with respect to the $\sigma$-field $\mathcal{F}^{k-1}$ generated by the history of play up to time $k-1$. Simply put, the noisy feedback $U_i^k$ is a conditionally unbiased estimator of $u_i^k$ with respect to the history, which is a standing assumption when dealing with the convergence of learning dynamics in games. For noisy feedback in general, or equivalently $\xi_i^k$ being a generic random variable, the discussion will be carried out in a different context. In that case, a system state should be introduced, which accounts for the uncertainty in the environment, and the learning problem becomes Nash equilibrium seeking in stochastic games (see \Cref{def:markov}). For more detailed discussions, we refer the reader to another paper in this special issue \cite{lei_csm}.     

The perfect global feedback is of limited use in practice when designing learning algorithms, as the global information is difficult or even impossible to acquire for individuals in large-scale network systems. For example, in distributed or decentralized learning over heterogeneous networks, players may have no access to others' utilities due to physical limitations. Therefore, we are interested in the scenario where players only have direct or indirect access to their own utilities as well as their neighbors', and hence players' feedback can be dependent on the topological structure of the underlying network that connects them.
 
Consider a repeated game over a  graph $\mathcal{G}:=(\mathcal{N},\mathcal{E})$, where $\mathcal{N}=\{1,2,\ldots,N\}$ is the set of nodes, representing the players in the game, who are connected via the edges in $\mathcal{E}=\{(i,j)|i,j\text{ are connected}\}$. To simplify the exposition, we assume that the graph is undirected.  Note that the direction of the edges does not affect our discussion as long as the neighborhood is properly defined. For example, in a directed graph, when in-neighbors or out-neighbors specify to which player(s) the player in question can pass information, then the following characterizations of feedback structures still apply. For a more comprehensive treatment of games over networks, we refer the reader to \cite{jackson15network_games}.

Each player is allowed to exchange payoff feedback with its neighbors through the edges and observe  their actions during the repeated play, whereas the information regarding the rest is hidden from him. In this case, the  feedback structure for player $i$ is  $$I_i^k=\{\{u_j^{1:k}\}_{j\in \{i\}\cup\mathcal{N}(i)}, \{a_j^{1:k}\}_{j\in \{i\}\cup\mathcal{N}(i) }\},\quad \mathcal{N}(i):=\{j|(i,j)\in \mathcal{E}\}.$$ Note that the player's  feedback regarding the payoffs and actions may not be consistent. For example, in a multi-agent robotic system where only the sensors network is effective, each agent can only observe its neighbors' movements through sensors. In this case, without any information of others' utilities, the information feedback of agent $i$ reduces to $I_i^k=\{\{u_i^{1:k}\}, \{a_j^{1:k}\}_{j\in\{i\}\cup \mathcal{N}(i)}\}$. To sum up, if the players can only receive feedback from their neighbors, then players'  feedback structures are related to the underlying topology,  leading to what is referred to as the \textit{local feedback}. In accordance with this, the extreme case of local  feedback is one where the player is isolated in the network, and no information other than its own payoff feedback and actions is available to it. We refer to this extreme case as \textit{individual  feedback}, which is a typical information feedback considered in fully decentralized learning and will be further elaborated on when discussing specific learning dynamics later in this section.     

In addition to the refinements from the spatial side, we can also consider  feedback with various temporal structures. If the player has perfect recall of previous plays, the resulting feedback is said to be \textit{perfect}, and those we have introduced above all fall within this class. Otherwise, players have access to \textit{imperfect feedback}, and we discuss two common cases of imperfect information feedback in the following, namely windowed and delayed feedback.

For the sake of simplicity, we use perfect  feedback $I_i^k=\{u_i^{1:k},a_i^{i:k}\}$ as a baseline to illustrate that different missing parts of $I_i^k$ lead to different kinds of imperfect feedback. If the head of  $u_i^{1:k}$ and/or  $a_i^{1:k}$ is not available to the player, that is, there exists a window $0<m<k$ such that the player only recalls $u_i^{(k-m):k}, a_i^{(k-m):k}$, then the corresponding feedback ${I}_i^{(k-m):k}=\{u_i^{(k-m):k}, a_i^{(k-m):k}\}$ is referred to as the windowed feedback with a window size $m$. Similarly, if the tail of $u_i^{1:k}$ and/or  $a_i^{1:k}$ is not available, that is, the player only recalls $u_i^{1:(k-m)}, a_i^{1:(k-m)}$, then the imperfect information feedback is ${I}_i^{1:(k-m)}=\{u_i^{1:(k-m)}, a_i^{1:(k-m)}\}$, which is called $m$-step delayed feedback.

\begin{figure}
    \centering
    \includegraphics[width=0.7\textwidth]{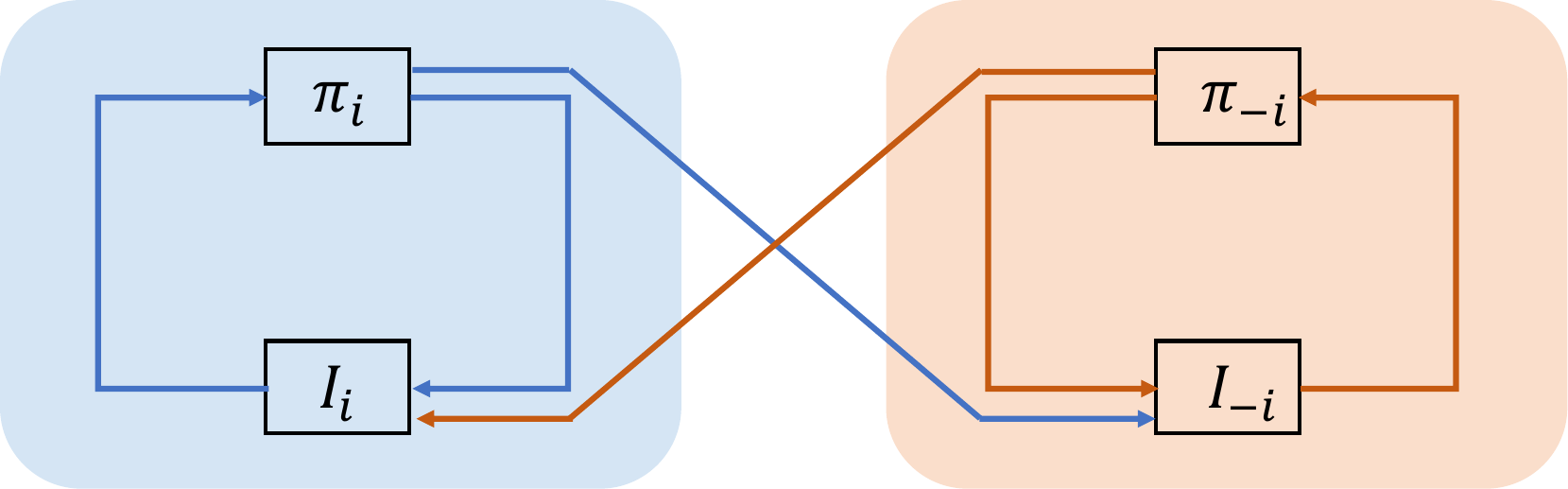}
    \caption{Player's strategy learning with the corresponding feedback. Under the global or the local feedback structure, players' learning processes are coupled, as their feedback is influenced by their opponents' moves. By contrast, players learn to play the game independently under the individual feedback.       }
    \label{fig:feedback}
\end{figure}

For learning in games, each player learns to select actions by updating the strategy based on the available feedback at each round. To describe this in mathematical terms, let $F_i^k$ the strategy learning policy of player $i$. The learning policy produces a new strategy $\pi_i^{k+1}$ for the next play according to
\begin{align}\label{eq:strategy_update}
    \pi_i^{k+1}=(1-\lambda_i^k)\pi_i^k+\lambda_i^k F_i^k(I_i^k),
\end{align}
where $\lambda_i^k$ is the learning rate, indicating the player's capabilities of information retrieval.  Different feedback structures lead to different learning dynamics in repeated games. Under the global or the local  feedback structure, each player's feedback is influenced by its opponents' actions and/or payoffs, which makes the players' learning processes coupled, as shown in \Cref{fig:feedback}.

In the case of fully decentralized learning under individual information feedback, players learn to play the game independently, and such a learning process is said to be uncoupled. Uncoupled learning processes are of great significance in both theoretical studies \cite{hart03uncoupled} and practical applications. Theoretically, learning with such limited information feedback is much more transferable in the sense that learning algorithms under this feedback also apply to online optimization problems, where the online decision-making process is viewed as a repeated game played between a player and the nature \cite{shai_online}. 

Considering its theoretical importance, we focus on learning with individual feedback in the sequel, and we refer the reader to \cite{marden14review_young} for a survey on learning methods under other kinds of feedback. We first present reinforcement learning for finite games, where the learning algorithms are characterized into two main classes, due to their distinct nature in exploration. Then, we proceed to gradient play for infinite games, and elaborate on its connection to reinforcement learning. The convergence results of presented algorithms are discussed in \Cref{sec:conver} based on stochastic approximation \cite{Borkar:2009ts,benaim05SADI} and Lyapunov stability theory.
%Despite their different applications, we demonstrate that the two are closely connected in the sense that reinforcement learning can be interpreted as a gradient play in finite games, while gradients of utility functions in gradient play serves as a performance evaluation in reinforcement learning.

\subsection{Reinforcement Learning}
Reinforcement learning has been studied in many disciplines and has become a catch-all term for learning in sequential decision making processes where the players' future choices of actions are shaped by the feedback. In general, reinforcement learning consists of two functions, one of which is the \textit{score function}, evaluating the performance of actions, and the other one is the \textit{choice mapping}, determining the next move. Note that in the machine learning literature \cite{Sutton:2018wc}, the score function and the choice mapping are also called the critic and the actor, respectively.  Different score functions and choice mappings lead to different reinforcement learning algorithms. We first provide a generic description of the score function and choice mapping in reinforcement learning from a dynamic system viewpoint, and then we give a characterization of various reinforcement learning algorithms based on different natures in choice mappings. Finally, at the end of this subsection, relations among introduced reinforcement learning algorithms are discussed.

To begin with, we show how the score function can be constructed using the information feedback recursively. Since the player has no direct access to its utility function in this case, it can construct an estimator $\hat{\mathbf{u}}_i^k\in \R^{|\mathcal{A}_i|}$ based on $I_i^k$ to evaluate actions $a\in \mathcal{A}_i$. By using this estimator, the player can compare its actions and choose the one that can achieve higher payoffs in the next round. In mathematical terms, the estimator (score function) is given by the following discrete-time dynamical system
\begin{align}
 \hat{\mathbf{u}}_i^{k+1} = (1-\mu_i^k)\hat{\mathbf{u}}_i^k +\mu_i^k G^k_i(\pi^k_i, \hat{\mathbf{u}}_i^k, U_i^k, a_i^k),\label{eq:utility}
\end{align}
where $G_i^k:\Delta(\mathcal{A}_i)\times\R^{|\mathcal{A}_i|}\times \R\times \mathcal{A}_i\rightarrow \R^{|\mathcal{A}_i|}$ is the learning policy for utility learning, $\pi_i^k$ is the policy employed at time $k$, and $\mu_i^k$ is the learning rate. Based on the score function, the player can modify its strategy accordingly in the sense that better actions shall be played more frequently in the future. With slight abuse of notations, the strategy update is 
\begin{align}
 \pi_i^{k+1}=(1-\lambda_i^k)\pi_i^k+\lambda_i^k F_i^k(\pi_i^k,\hat{\mathbf{u}}_i^{k+1}, U_i^k, a_i^k),\label{eq:strategy}
 \end{align}
 where $F_i^k:\Delta(\mathcal{A}_i)\times\R^{|\mathcal{A}_i|}\times \R\times \mathcal{A}_i\rightarrow \Delta(\mathcal{A}_i)$ is the learning policy for strategy learning, yielding a new policy for the next play. Compared with \eqref{eq:strategy_update}, the above discrete-time systems \eqref{eq:utility} \eqref{eq:strategy}  explicitly show how the feedback shapes the player's future play. According to \eqref{eq:utility}, the player recursively updates its estimate of the utility function based on the feedback it receives after playing $\pi_i^k$, and then the player determines its move in the next round, following \eqref{eq:strategy}.  Intuitively, we can view $(\pi_i^k, \hat{\mathbf{u}}_i^{k+1})$ as the information extracted from $I_i^k$ for updating the player's strategy.

In reinforcement learning, the choice mapping plays an important role in achieving the balance between exploitation and exploration. On one hand, the player would like to choose the best action that is supposed to incur the highest payoff based on the score function. However, this pure exploitation oftentimes leads to myopic behaviors, as the score function may return a poor estimate of the utility function at the beginning of the learning process. Hence, to gather more information for a better estimator, the player also needs some experimental moves for exploration, where suboptimal actions are implemented. To sum up, the trade-off between exploitation and exploration is of vital importance to the success of reinforcement learning, and it depends on the construction of the choice mapping. Different choice mappings result in different reinforcement learning algorithms. Based on their distinct natures in exploration, the algorithms can be categorized into two main classes: \textit{exploitative reinforcement learning} and \textit{exploratory reinforcement learning}.  

Recall that in the strategy learning \eqref{eq:strategy}, the next strategy produced by the corresponding choice mapping is   
$$ \pi_i^{k+1}=(1-\lambda_i^k)\pi_i^k+\lambda_i^k F_i^k(\pi_i^k,\hat{\mathbf{u}}_i^{k+1}, U_i^k, a_i^k),$$ 
where $(1-\lambda_i^k)\pi_i^k$ is referred to as the \textit{cognitive inertia} or simply \textit{inertia}, describing the player's tendency to repeat previous choices independently of the outcome. When determining its next move $\pi_i^{k+1}$, the player takes into account both its previous strategy $\pi_i^{k}$ and the increment update using the strategy learning policy $F_i^k$. Therefore, players' exploration at $(k+1)$-th round stems either from this inertia or the strategy learning policy $F_i^k$. The former is called \textit{passive exploration}, as it relies on the player's tendency to repeat previous choices, while the latter one is referred to as \textit{active exploration}, as the player deliberately tries actions based on what he has learned from previous plays. 

As the new strategy is a convex combination of the inertia term $\pi_i^k$ and the learned incremental update $F_i^k(\pi_i^k,\hat{\mathbf{u}}_i^{k+1}, U_i^k, a_i^k)$, there is no clear-cut boundary between passive and active exploration. In fact, reinforcement learning is a continuum of learning algorithms. In the following, we illustrate such a continuum by three prominent learning schemes. The first one is the best response dynamics, located on the left endpoint, which is an example of  exploitative reinforcement learning. Solely relying on the inertia for passive exploration, the best response dynamics adopts a purely exploitative learning policy: the best response mapping in \eqref{eq:br_nash}.   On the contrary to the exploitative one, we present dual averaging as an example of exploratory reinforcement learning, which only leverages the learning policy for exploring suboptimal actions without any cognitive inertia. In between, there lies the smoothed best response dynamics, where both the inertia and the strategy learning policy come into play for achieving the balance between exploration and exploitation.

\subsubsection{Exploitative Reinforcement Learning}

For exploitative reinforcement learning, the strategy learning policy always outputs the best strategy based on the score function, which can be viewed as a natural extension of the best response idea in the context of Nash equilibrium \eqref{eq:br_nash}. In the repeated play scenario, given the opponent's strategy at the $k$-th round, $\pi_{-i}^k$, from player $i$'s standpoint, the best it can do is to choose the best response  $BR_i(\pi_{-i}^{k}):=\argmax_{\pi\in \Delta(\mathcal{A}_i)}\{\left\langle \pi, \mathbf{u}_i(\pi_{-i}^k)\right\rangle\}$, which is purely exploitative. In this case, the strategy learning scheme becomes
\begin{align}\label{eq:br_dis}
	\pi_i^{k+1}\in (1-\lambda_i^k)\pi_i^k+\lambda_i^k BR_i(\pi_{-i}^{k}).
\end{align}
In general, the best response mapping is a point-to-set mapping, and to analyze the associated learning dynamics, differential inclusion theory \cite{benaim05SADI} is needed, which make the convergence analysis more involved as discussed in \Cref{sec:nash_lyap}. 

Under the noisy feedback $I_i^k=\{U_i^{1:k},a_i^{1:k}\}$, the score function of player $i$ is the estimated utility $\hat{\mathbf{u}}_i^k$, which is updated according to the following moving average scheme \cite{leslie03sbr_multi}
\begin{align}\label{eq:average}
	\hat{\mathbf{u}}_i^{k+1}(a)=(1-\mu_i^k)\hat{\mathbf{u}}_i^k(a)+\mu_i^k\frac{\mathbbm{1}_{\{a=a_i^k\}}}{\pi_i^k(a)}U_i^k,\quad a\in \mathcal{A}_i,
\end{align} 
where $\mathbbm{1}_{\{\cdot\}}$ is an indicator function. Note that in \eqref{eq:average}, the importance sampling technique, which is common in bandit algorithms [15], is utilized to construct an unbiased estimator of $\mathbf{u}_i(\pi_{-i}^k)$. To see this,  define a vector $\hat{\mathbf{U}}_i^k\in\R^{|\mathcal{A}_i|}$, whose $a$-th entry is  $\hat{\mathbf{U}}_i^k(a):=\mathbbm{1}_{\{a=a_i^k\}}U_i^k/\pi_i^k(a)$; and we then obtain $\mathbb{E}[\hat{\mathbf{U}}_i^k(a)|\mathcal{F}^{k-1}]=u_i(a,\pi_{-i}^k)$. Hence, \eqref{eq:average} can be rewritten as 
\begin{align}\label{eq:moving_aver}
	\hat{\mathbf{u}}_i^{k+1}=(1-\mu_i^k)\hat{\mathbf{u}}_i^k+\mu_i^k\hat{\mathbf{U}}_i^k,
\end{align}
and $\hat{\mathbf{u}}_i^{k+1}(a)$ gives the averaged payoff incurred by $a$ in the first $k$ rounds. This importance sampling technique can be viewed as compensating for the fact that actions played with a low probability do not receive frequent updates of the corresponding estimates, so that when they are played, any estimation error $U_i^k-\hat{\mathbf{u}}_i^k(a_i^k)$ must have greater influence on the estimated value than if frequent updates occur. We refer the reader to \cite{shai_online,fuden_learning} for more details on importance sampling and its use in learning processes.

With a slight abuse of the notation of best response mapping in \eqref{eq:br_mapping}, we define the corresponding best response under the noisy feedback as 
\begin{align}
	BR_i(\hat{\mathbf{u}}_i^k):=\argmax_{\pi\in \Delta(\mathcal{A}_i)}\{\left\langle \pi,\hat{\mathbf{u}}_i^k\right\rangle\}.
\end{align}
Then, we obtain the following strategy learning scheme \cite{fuden_learning}
\begin{align}\label{eq:br}
\pi_i^{k+1}\in (1-\lambda_i^k)\pi_i^k+\lambda_i^k BR_i(\hat{\mathbf{u}}_i^k).
\end{align}
The resulting dynamical system under the noisy feedback is a coupled system as shown below 
\begin{align}
\begin{aligned}\label{eq:br_noisy}
    \hat{\mathbf{u}}_i^{k+1}&=(1-\mu_i^k)\hat{\mathbf{u}}_i^k+\mu_i^k\hat{\mathbf{U}}_i^k,\\
    \pi_i^{k+1}&\in (1-\lambda_i^k)\pi_i^k+\lambda_i^k BR_i(\hat{\mathbf{u}}_i^k).
\end{aligned}\tag{\text{BR-d}}
\end{align}
Originally proposed as a computational method for Nash equilibrium seeking \cite{harris98con_fp,leslie03sbr_multi}, the best response dynamics \eqref{eq:br_noisy} is directly built upon the best response idea and has been widely applied to evolutionary game problems \cite{hofbauer03evolu_dyna}. One prominent example of best response dynamics is fictitious play \cite{brown1951iterative}, where a player's empirical play follows \eqref{eq:br_noisy}; and more details are included in \Cref{side:fp}. As shown above, best response dynamics adopts passive exploration, and the best response mapping $BR_i(\cdot)$ encourages greedy actions that might be myopic. As a result, exploitative reinforcement learning may fail to converge \cite{fuden_learning,krishna98br_fail}.

\subsubsection{Exploratory Reinforcement Learning}
In contrast to the inertia-based passive exploration in \eqref{eq:br_noisy}, dual averaging, as introduced in this subsection, only relies on the strategy learning policy $F_i^k$ for exploring suboptimal actions, in order to avoid myopic behaviors due to the poor estimates of the utility function. In dual averaging, given the player's utility vector $\mathbf{u}_i$, the strategy learning policy is a regularized best response \cite{PM18cont}, defined as  
 \begin{align}\label{eq:mirror}
QR^\epsilon(\mathbf{u}_i):=\argmax_{\pi_i\in\Delta(\mathcal{A}_i)}\{\left\langle\pi_i, \mathbf{u}_i \right\rangle- \epsilon h(\pi_i)\}, 
\end{align}
where $h(\cdot)$ is a penalty function or regularizer and $\epsilon$ is the regularization parameter. According to \cite{PM16lyapnov}, a proper regularizer $h(\cdot)$ defined on the probability simplex should be continuous over the simplex and smooth on the relative interior of every face of the simplex. Besides, $h$ should be a strongly convex function, and these assumptions ensure that $QR^\epsilon(\cdot)$ always returns a unique maximizer. The mapping $QR^\epsilon$ is referred to as a quantal response mapping \cite{mckelvey95q}, which allows players to choose suboptimal actions with positive probability. To see how this regularization contributes to active exploration, consider the entropy regularizer $h(x)=\sum_{x_i}x_i\log x_i$. In this case, $QR^\epsilon$ is  
\begin{align}\label{eq:bg}
	QR^\epsilon(\mathbf{u}_i)(a):=\frac{\exp(\frac{1}{\epsilon}u_i(a,\pi_{-i}))}{\sum_{a'\in \mathcal{A}_i}\exp(\frac{1}{\epsilon}u_i(a',\pi_{-i}))}, \quad a\in\mathcal{A}_i,
\end{align} 
which is also known as the Boltzmann-Gibbs strategy mapping 
\cite{zhu10heter} or the soft-max function parameterized by $\epsilon>0$. On the one hand, the Boltzmann-Gibbs mapping produces a strategy that assigns more weight to the actions leading to higher payoffs, that is, the larger $\mathbf{u}_i(a)=u_i(a,\pi_{-i})$ is, the larger $QR^\epsilon(\mathbf{u}_i)(a)$ becomes. On the other hand, it always retains positive probabilities for every action, when $\epsilon>0$. Note that $QR^\epsilon$ can induce different levels of exploration by adjusting the parameter $\epsilon$. When $\epsilon$ tends to 0, the strategy \eqref{eq:bg} simply returns the action that yields the highest payoff, implying that $QR^\epsilon$ reduces to the best response mapping $BR_i(\cdot)$ in \eqref{eq:br_mapping}.  As $\epsilon$ gets larger, $1/\epsilon$ tends to 0, and the strategy does not distinguish among actions, leading to equal weights for all actions.

Similar to the previous argument, with the noisy feedback, we replace $\mathbf{u}_i$ by the estimator $\hat{\mathbf{u}}_i^k$, and the definition of  quantal response mapping is then modified accordingly as
\begin{align*}
	QR^\epsilon(\hat{\mathbf{u}}_i^k)(a):=\frac{\exp(\frac{1}{\epsilon}\hat{\mathbf{u}}_i^k(a))}{\sum_{a'\in \mathcal{A}_i}\exp(\frac{1}{\epsilon}\hat{\mathbf{u}}_i^k(a'))}, \quad a\in\mathcal{A}_i.
\end{align*}
Due to the active exploration brought up by $QR^\epsilon$, we can consider an inertia-free reinforcement learning scheme, where the choice map is simply the strategy learning policy $QR^\epsilon$.  The corresponding strategy learning scheme is then as
\begin{align*}
	\pi_i^{k+1}=QR^\epsilon(\hat{\mathbf{u}}_i^{k+1}),
\end{align*}   
where the score function $\hat{\mathbf{u}}_i^k$ is updated according to the following \cite{nesterov09da}
\begin{align}\label{eq:cumulative}
    \hat{\mathbf{u}}_i^{k+1}=\hat{\mathbf{u}}_i^k+\mu_i^k \hat{\mathbf{U}}_i^k.
\end{align}

To recap, the learning algorithm operates in the following fashion: at each time $k$, an unbiased estimator $\hat{\mathbf{U}}_i^k$ is constructed as introduced in \eqref{eq:average}, using importance sampling, and the score function is updated according to \eqref{eq:cumulative}. Then, the next strategy is produced by the mapping $QR^\epsilon$, acting on the score function $\hat{\mathbf{u}}_i^{k+1}$, as shown below
\begin{align}\label{eq:da}
	\begin{aligned}
		&\hat{\mathbf{u}}_i^{k+1}=\hat{\mathbf{u}}_i^k+\mu_i^k \hat{\mathbf{U}}_i^k,\\
		&\pi_i^{k+1}=QR^\epsilon(\hat{\mathbf{u}}_i^{k+1}),
	\end{aligned}\tag{DA-d}
\end{align} 
\eqref{eq:da} is also referred to as dual averaging, pioneered by Nesterov \cite{nesterov09da}, which was originally  proposed as a variant of gradient methods for solving convex programming problems.  We elucidate the term ``dual averaging'' later when we discuss the relation between dual averaging and gradient play, where we demonstrate that \eqref{eq:da} can be viewed as a gradient-based algorithm in finite games with $\hat{\mathbf{u}}_i^k$ being the gradient. Finally, as a remark, we note that in \eqref{eq:da}, the score function is updated in a manner different than in best response dynamics \eqref{eq:br_noisy}. However, this is merely a matter of presentation, and by selecting a proper $\epsilon$, the moving averaging scheme \eqref{eq:moving_aver} is essentially the same as the discounted accumulation \eqref{eq:cumulative}, for which we refer the reader to \cite{nesterov04book,nesterov09da}. By adopting the discounted accumulation \eqref{eq:cumulative}, we later can draw a connection between dual averaging and gradient play.

Apparently, the discrete-time system \eqref{eq:da} does not depict how $\pi_i(t)$ evolves in $\Delta(\mathcal{A}_i)$, and it is not straightforward to tell how those good actions bringing up higher payoffs are ``reinforced'' in the sense that probabilities of choosing them are increasing as the learning process proceeds. In \Cref{side:rd}, we present that when choosing entropy regularization, \eqref{eq:da} is equivalent to the replicator dynamics, one of the well-known evolutionary dynamics \cite{maynard73logic,sandholm10evolu_game,cressman14replicator}, which explicitly displays a gradual adjustment of strategies based on the quality of each action. Meanwhile, with an example of population games, we show that this connection brings learning in games to the broader context of evolutionary game theory \cite{sandholm10evolu_game, hofbauer03evolu_dyna}.  

As we have mentioned, reinforcement learning is a continuum of learning algorithms, and the best response dynamics \eqref{eq:br_noisy} and dual averaging \eqref{eq:da} are the two endpoints of the continuum. Naturally, we can consider reinforcement learning methods with a blend of both passive and  active exploration, where the exploration stems from both the inertia term and the strategy learning policy, as we present in the following.

Instead of choosing actions greedily, we replace the best response $BR_i(\cdot)$ in \eqref{eq:br} by $QR^\epsilon(\cdot)$, the quantal response  for active exploration, and then we obtain the following strategy learning scheme \cite{fuden_learning}
\begin{align*}
    \pi_i^{k+1}=(1-\lambda_i^k)\pi_i^k+\lambda_i^k QR^\epsilon(\hat{\mathbf{u}}_i^k).
\end{align*}
Similar to the best response dynamics in \eqref{eq:br_noisy}, if utility learning follows the moving average scheme in \eqref{eq:average}, the resulting reinforcement learning has the following discrete-time learning dynamics 
\begin{align}\label{eq:sbr}
    \begin{aligned}
\hat{\mathbf{u}}_i^{k+1}&=(1-\mu_i^k)\hat{\mathbf{u}}_i^k+\mu_i^k\hat{\mathbf{U}}_i^k,\\
	\pi_i^{k+1}&=(1-\lambda_i^k)\pi_i^k+\lambda_i^k QR^\epsilon(\hat{\mathbf{u}}_i^k).
\end{aligned}\tag{\text{SBR-d}}
\end{align}
Considering its similarity to best response dynamics, \eqref{eq:sbr} is referred to as smoothed best response dynamics in the literature \cite{fuden_learning,leslie06sbr}. Specifically, if the entropy regularizer is adopted, the resulting learning process is called Boltzmann-Gibbs reinforcement learning \cite{zhu12heter} or entropic reinforcement learning, which has been extensively studied in the context of Markov decision processes \cite{neu2017unified}. 

\subsubsection{Relations among Reinforcement Learning Algorithms}

\begin{figure}
    \centering
    \includegraphics[width=0.6\textwidth]{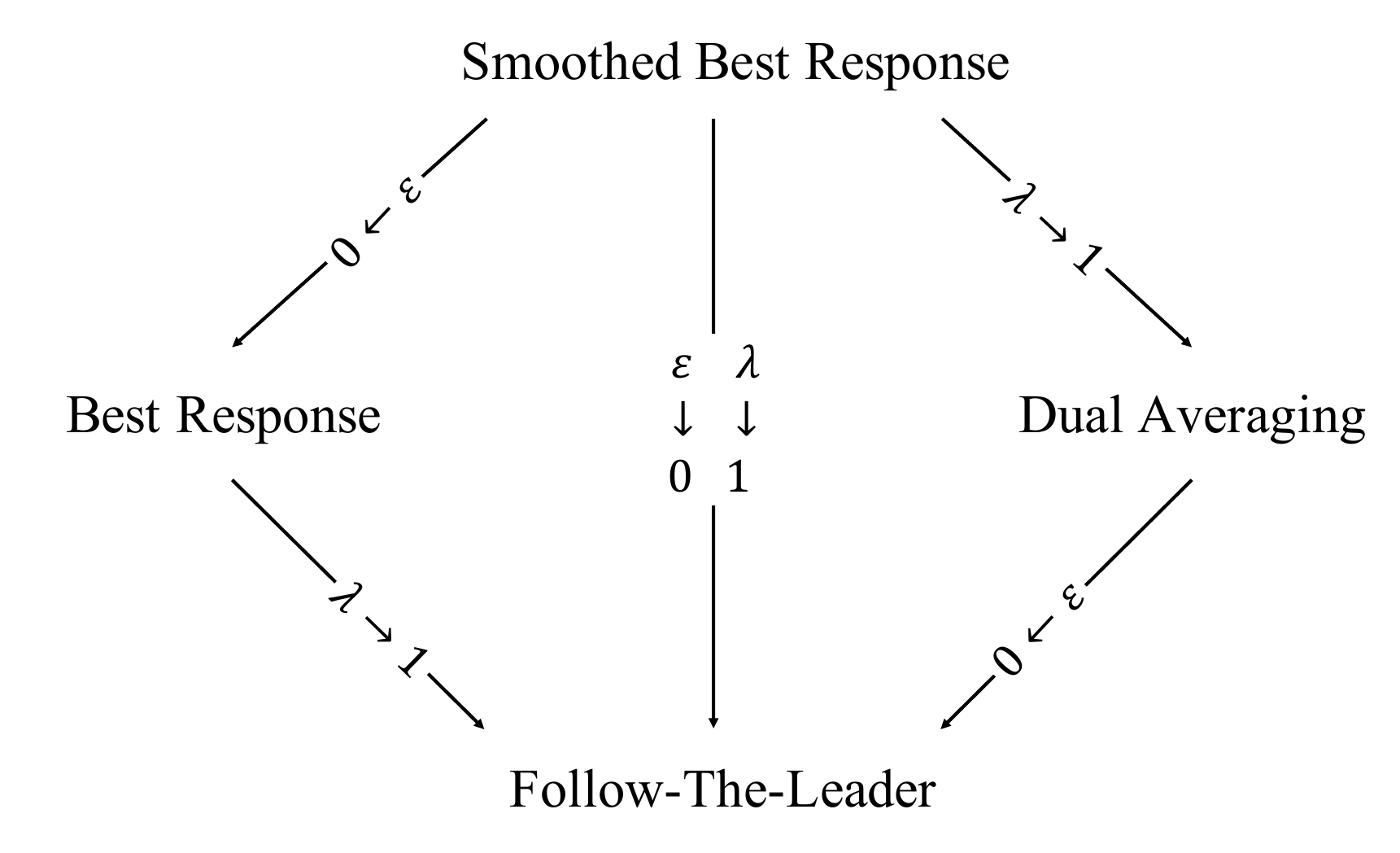}
    \caption{Relationships of reinforcement learning algorithms. For $0<\lambda<1$ and $\epsilon>0$, we obtain the exploratory reinforcement learning: smoothed best response dynamics \eqref{eq:sbr}, where exploration arises from both the inertia and the learning policy. If the active exploration vanishes as $\epsilon$ goes to zero, smoothed best response reduces to best response dynamics \eqref{eq:br_noisy}, an example of exploitative reinforcement learning. By contrast, we obtain dual averaging \eqref{eq:da_con}, if $\lambda$ tends to 1. Finally, if  $\epsilon$ goes to zero while $\lambda$ tends to 1, players always choose their actions greedily according to follow-the-leader policy. }
    \label{fig:algo}
\end{figure}  

Before wrapping up our presentation on reinforcement learning in finite games, we discuss the relations among the introduced learning algorithms. We reiterate that reinforcement learning corresponds to a continuum of learning algorithms, where one algorithm can be converted to the other by adjusting the learning rate $\lambda_i^k$ in strategy learning \eqref{eq:strategy_update} and/or the exploration parameter $\epsilon$, and a diagram of such conversion is presented in \Cref{fig:algo}. Our discussion will revolve around the learning rate $\lambda_i^k$ and the exploration parameter $\epsilon$. For simplicity, we suppress the subscript and the superscript of the learning rate and simply denote it by $\lambda$.  

We begin the discussion with the learning rate $\lambda$. Different from dual averaging \eqref{eq:da}, the best response dynamics \eqref{eq:br_noisy} and the smoothed best response dynamics \eqref{eq:sbr} are in fact actor-critic learning \cite{konda99two,leslie03sbr_multi,leslie05sbr_q} due to a positive learning rate $\lambda>0$. Under the actor-critic framework such as \eqref{eq:br_noisy}\eqref{eq:sbr}, the player maintains two recursive schemes for updating the estimated utility vector and the strategy, respectively.  The recursive schemes lead to coupled dynamical systems of $\hat{\mathbf{u}}_i^{k}$ and $\pi_i^k$. In contrast, even though dual averaging \eqref{eq:da} also consists of both updating schemes for estimated utility vector and the strategy, since the learning rate is zero, there is only one effective dynamical system: the one induced by the estimation of utility vector \eqref{eq:cumulative}. Another way to see the difference between actor-critic learning \eqref{eq:br_noisy}\eqref{eq:sbr} and dual averaging \eqref{eq:da} is through the corresponding continuous-time learning dynamics in \Cref{sec:con_dyna}. 

Even though \eqref{eq:da} is not an actor-critic learning, its trajectory is closely related to that of \eqref{eq:br_noisy}\eqref{eq:sbr}. Intuitively speaking, dual averaging only differs from the smoothed best response in that \eqref{eq:da} does not acquire an inertia term, as the learning rate is zero. Hence, $\pi_i^k$ in \eqref{eq:sbr} can be seen as the moving average of $QR^\epsilon(\hat{\mathbf{u}}_i^{k})$ in \eqref{eq:da}. Therefore, it is reasonable to expect that the time average of the trajectory produced by \eqref{eq:da} is related to the one produced by the smoothed best response. This intuition has been verified in \cite{horbauer09time,PM16lyapnov}, where it has been shown that the time averaged trajectory of \eqref{eq:da} follows \eqref{eq:sbr} with a  time-dependent perturbation $\epsilon(t)$.    

Apart from the difference in the learning rates, learning algorithms also display distinct asymptotic behavior due to the difference in the exploration parameter. The exploration parameter $\epsilon$ has less drastic consequence under \eqref{eq:da} than under the actor-critic learning $\eqref{eq:br_noisy}\eqref{eq:sbr}$. As observed in \cite{PM16lyapnov}, adding a positive $\epsilon$ is equivalent to rescaling the regularizer, that is, replacing $h(\cdot)$ with $\epsilon h(\cdot)$. As long as $\epsilon>0$, the regularization $\epsilon h(\cdot)$ is still proper (see \eqref{eq:mirror} and the following discussion).  This implies that even though the choice of $\epsilon$ affects the speed at which \eqref{eq:da} evolves, the qualitative results remain the same. We refer the reader to \cite{PM16lyapnov, PM18rieman} for a detailed discussion. When $\epsilon=0$, there is no exploration nor inertia for dual averaging, and in this case, players always choose their actions greedily according to the best response mapping
\begin{align}\label{eq:ftl}
    \pi_i^{k+1}=\argmax_{\pi\in \Delta(\mathcal{A}_i)}\{\langle \pi, \hat{\mathbf{u}}_i^{k}\rangle\}\tag{\text{FTL}},
\end{align}
where $\hat{\mathbf{u}}_i^{k}$ is the score function of player $i$, based on its history of play up to round $k$, and it can be updated following \eqref{eq:average} or \eqref{eq:cumulative}. In the online learning literature \cite{shai_online}, \eqref{eq:ftl} is known as \textit{follow-the-leader (FTL)} policy, which can also be obtained by eliminating the inertia term in the best response dynamics \eqref{eq:br_noisy}. Due to lack of exploration, \eqref{eq:ftl} is too aggressive and can be exploited by the adversary, resulting in a positive, non-diminishing regret \cite{shai_online}. The regret is a measure of the performance gap between the cumulative payoffs of current policy \eqref{eq:ftl} and that of the best policy in hindsight.     

The exploration parameter plays a more important role in the actor-critic learning which balances exploration and exploitation \cite{Sutton:2018wc}. The smoothed best response \eqref{eq:sbr}, which is a perturbed version of the best response, can only use the regularization $\epsilon h(\cdot)$ for encouraging active exploration. Thanks to the positive exploration parameter, the smoothed best response \eqref{eq:sbr} enjoys an $\epsilon$-no-regret property, a weak form of external consistency studied in \cite{benaim06SADI,horbauer09time}, which is desired in an adversarial environment \cite{shai_online}.  In contrast, the best response dynamics \eqref{eq:br_noisy}, due to the myopic nature of the best response mapping \eqref{eq:br_mapping}, does not possess  similar properties.

\subsection{Gradient Play}\label{sec:gp}
Heretofore, we have limited our discussions to learning processes in finite games, where the score function \eqref{eq:utility} and the choice mapping \eqref{eq:strategy} act on finite-dimensional vectors. For continuous-kernel games, it is not straightforward to extend reinforcement learning, since a suitable score function is required to evaluate a continuum of actions, and constructing such a score function can be very challenging.  Even though function approximators, such as linear \cite{geramifard13linear_approx,tao_multiRL} or nonlinear \cite{mnih2015DQN} ones can be of some help, we present here a mathematically more elegant way of leveraging the reinforcement idea based on gradients of utility functions. In other words, instead of seeking the maximizers, we seek for a better response by searching along the gradient direction. Such gradient-based learning algorithms, referred to as gradient play, are popular in a variety of multi-agent settings due to their versatility, ease of implementation, and dependence on local information. 

\begin{figure}
    \centering
    \includegraphics[width=0.5\textwidth]{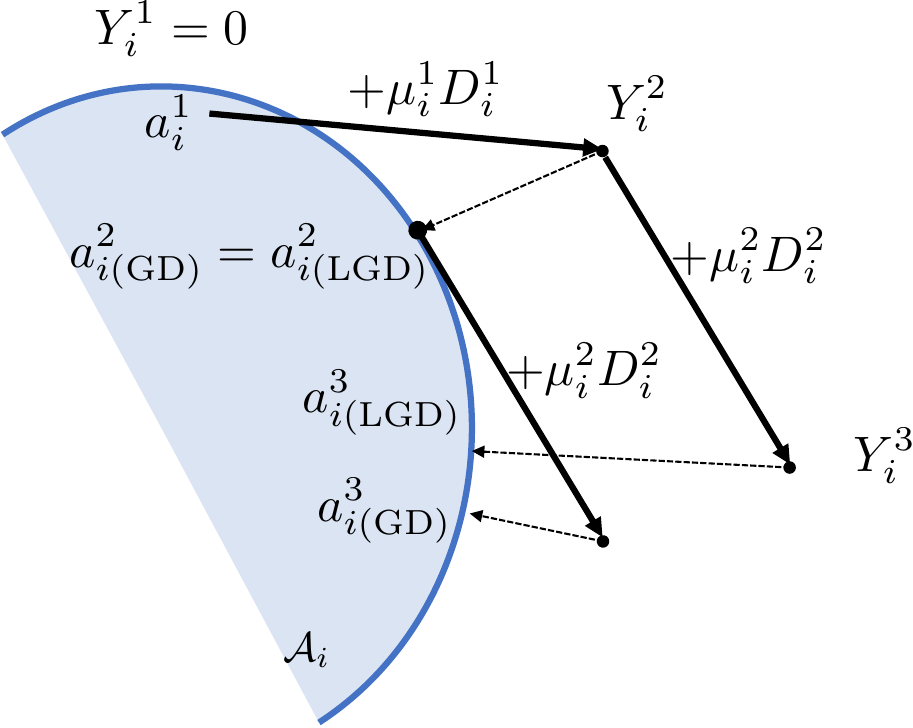}
    \caption{Illustration of the difference between \eqref{eq:pgd} and \eqref{eq:lgd}. $a_{i(\text{GD})}^k$ and $a_{i(\text{LGD})}^k$ denote, respectively, the iterates generated by \eqref{eq:pgd} and \eqref{eq:lgd}. \eqref{eq:lgd} first aggregates the gradient steps, and then projects the aggregation onto the primal space to generate a new gradient step.}
    \label{fig:lgd}
\end{figure}

For the sake of simplicity, we restrict our discussion to pure strategy Nash equilibrium in continuous games (see \eqref{eq:concave_ineq} for the definition and \eqref{eq:vi_ne} for its variational characterization), in order to avoid measure-theoretic issues when studying the mixed strategy case. We further assume that utilities are smooth functions and perfect feedback is available to players, implying that each player can compute the gradient of the utility function given current iterates: $D_i^k=\nabla_{a_i}u_i(a_i^k,a_{-i}^k)$. Even though the perfect feedback is assumed here, it is purely for the simplicity of exposition. It is viable for players to estimate the gradient based on the realized payoff under noisy individual feedback by simultaneous perturbation stochastic approximation \cite{spall97spsa,bravo18bandit_learning}. Based on this gradient, players update their actions according to the following 
\begin{align}\label{eq:pgd}
    a_i^{k+1}&=\operatorname{proj}_{\mathcal{A}_i}[a_i^k+\mu_{i}^k D_i^k ],\nonumber\\
    &:=\argmin_{a\in \mathcal{A}_i}\{\|a_i^k+\mu_{i}^k D_i^k-a\|_2^2\}\tag{\text{GD}}
\end{align}
where $\operatorname{proj}_{\mathcal{A}_i}(\cdot)$ is the Euclidean projection operator, and \eqref{eq:pgd} is called online gradient descent or projected gradient descent \cite{nesterov04book}. One extensively studied variant of \eqref{eq:pgd} \cite{nesterov04book,da_lin} is 
\begin{align}\label{eq:lgd}
    \begin{aligned}
    &Y_i^{k+1}=Y_i^k+\mu_i^k D_i^k,\\
    & a_i^{k+1}=\operatorname{proj}_{\mathcal{A}_i}(Y_i^{k+1}), 
    \end{aligned}\tag{\text{LGD}}
\end{align}
where $Y_i^{k}$ is an auxiliary variable that aggregates the gradient steps. Such an algorithm is referred to as the lazy gradient descent (LGD) \cite{nesterov09da}, since the algorithm aggregates the gradient steps ``lazily'', without transporting them to the action space as \eqref{eq:pgd} does. We illustrate the difference between the two algorithms in \Cref{fig:lgd}. We note that both based on the gradient descent idea, \eqref{eq:lgd} and \eqref{eq:pgd} share the same asymptotic behavior \cite{shai_online}, and the two coincide when $\mathcal{A}_i$ is an affine subspace of $\R^n$. 

Different from a purely primal-based algorithm, such as \eqref{eq:pgd}, where the trajectory of the algorithm only evolves in the primal space (the action space),  \eqref{eq:lgd} is a primal-dual scheme, and the interplay between primal variables, actions $a_i^k$, and the dual, gradients $D_i(\mathbf{a}^k)$, is of great significance. The main idea of \eqref{eq:lgd} is as follows. At the $k$-th round, each player computes the gradient $D_i(\mathbf{a}^k)$ based on the knowledge of utility functions and observations of the opponent's move. Subsequently, they take a step along this gradient in the dual space (where gradients live) and they ``mirror'' the output back to the primal space (the action space), using the Euclidean projection.

Gradient based learning algorithms are further investigated in another paper of this special issue in the context of generalized Nash equilibrium seeking \cite{pavel_csm}. In the following, we present a generalization of \eqref{eq:lgd}:  mirror descent \cite{nesterov09da,da_lin}. Starting with some arbitrary initialization $Y_i^1$, the mirror descent scheme can be described via the recursion
\begin{align}\label{eq:omd}
\begin{aligned}
       &Y_i^{k+1}=Y_i^k+\mu_i^k D_i(\mathbf{a}^k),\\
    & a_i^{k+1} = QR^\epsilon(Y_i^{k+1}), 
\end{aligned}\tag{\text{MD}}
\end{align}
where $QR^\epsilon$ is the quantal response mapping in the context of the continuous game, defined  as $$QR^\epsilon(Y)=\argmax_{a\in\mathcal{A}_i}\{\langle Y,a\rangle-\epsilon h(a)\}.$$ When we choose the Euclidean norm as the regularizer, that is, $h(x)=\frac{1}{2}\|x\|_2^2$ and $\epsilon=1$, $QR^\epsilon$ reduces to the projection operator $\operatorname{proj}_{\mathcal{A}_i}$. Geometrically, the gradient search step is performed in the dual space, and then the primal update is produced by the mapping $QR^\epsilon$. Since $QR^\epsilon$ ``mirrors'' the gradient update in the dual space back to the primal space, it is also referred to as the mirror map in the online optimization literature \cite{shai_online}. 

\subsubsection{Mirror Descent as Reinforcement Learning in Continuous Games}
Mirror descent \eqref{eq:omd} and the reinforcement learning \eqref{eq:da} share the same choice map, and they are closely connected. We demonstrate in the following that as a gradient-based algorithm, mirror descent can also be cast as a reinforcement learning scheme in continuous games, with $Y_i^k$ being the ``score function''.  

To evaluate a certain action $a\in \mathcal{A}_i$ at time $k$, consider $\sum_{\tau=1}^k u_i(a,a_{-i}^\tau)$, the counterfactual outcome had player $i$ implemented $a$ all the time in the past. The higher the sum is, the better action is $a$, since it could have brought up higher payoffs. Hence, the player can choose the next action that is optimal in hindsight:
\begin{align}\label{eq:ftrl}
    a_i^{k+1}=\argmax_{a\in\mathcal{A}_i}\{\sum_{\tau=1}^k u_i(a,a_{-i}^\tau)-\epsilon h(a)\},\tag{\text{FTRL}}
\end{align}
 where $\epsilon h(\cdot)$ is the regularization introduced in \eqref{eq:mirror}, encouraging exploration in the learning process. Based on the optimality in  hindsight, this action selection \eqref{eq:ftrl} is known as \textit{follow-the-regularized-leader} (FTRL) \cite{mertikopoulos18cycles}. Moreover, if $u_i$ is well-behaved in the sense that it can be approximated by the first-order Taylor expansion, that is, $u_i(a,a_{-i}^\tau)\approx u_i(a_i^\tau,a_{-i}^\tau)+\langle D_i(\mathbf{a}^\tau), a-a_i^\tau\rangle$, then \eqref{eq:ftrl} is equivalent to 
 \begin{align*}
     a_i^{k+1}&=\argmax_{a\in\mathcal{A}_i}\{\sum_{\tau=1}^k \langle D_i(\mathbf{a}^\tau), a\rangle -\epsilon h(a)\}\\
     &=\argmax_{a\in\mathcal{A}_i}\{ \langle \sum_{\tau=1}^kD_i(\mathbf{a}^\tau), a\rangle -\epsilon h(a)\}\\
     &=QR^\epsilon(\sum_{\tau=1}^kD_i(\mathbf{a}^\tau)),
 \end{align*}
which is exactly the mirror descent scheme in \eqref{eq:omd}, despite using an auxiliary variable $Y_i^k$ to aggregate these gradients weighted by the learning rates $\mu_i^k$. In other words, by the first-order expansion, the sum of gradients living in the dual space serves a linear functional for evaluating the quality of the actions.  Hence, the sum or equivalently $Y_i^k$ can be treated as a ``score function'', based on which the mirror map outputs a better action in hindsight, yielding a reinforcement procedure.        
\subsubsection{Reinforcement Learning as Mirror Descent in Finite Games}
In the above discussion, we interpreted the mirror descent scheme \eqref{eq:omd} as a ``reinforcement learning'' in continuous games. In this subsection, we further show that the idea of mirror descent can also be employed in finite games, and the resulting learning dynamics is in fact the exploratory reinforcement learning scheme \eqref{eq:da}.

\begin{figure}
    \centering
    \includegraphics[width=0.6\textwidth]{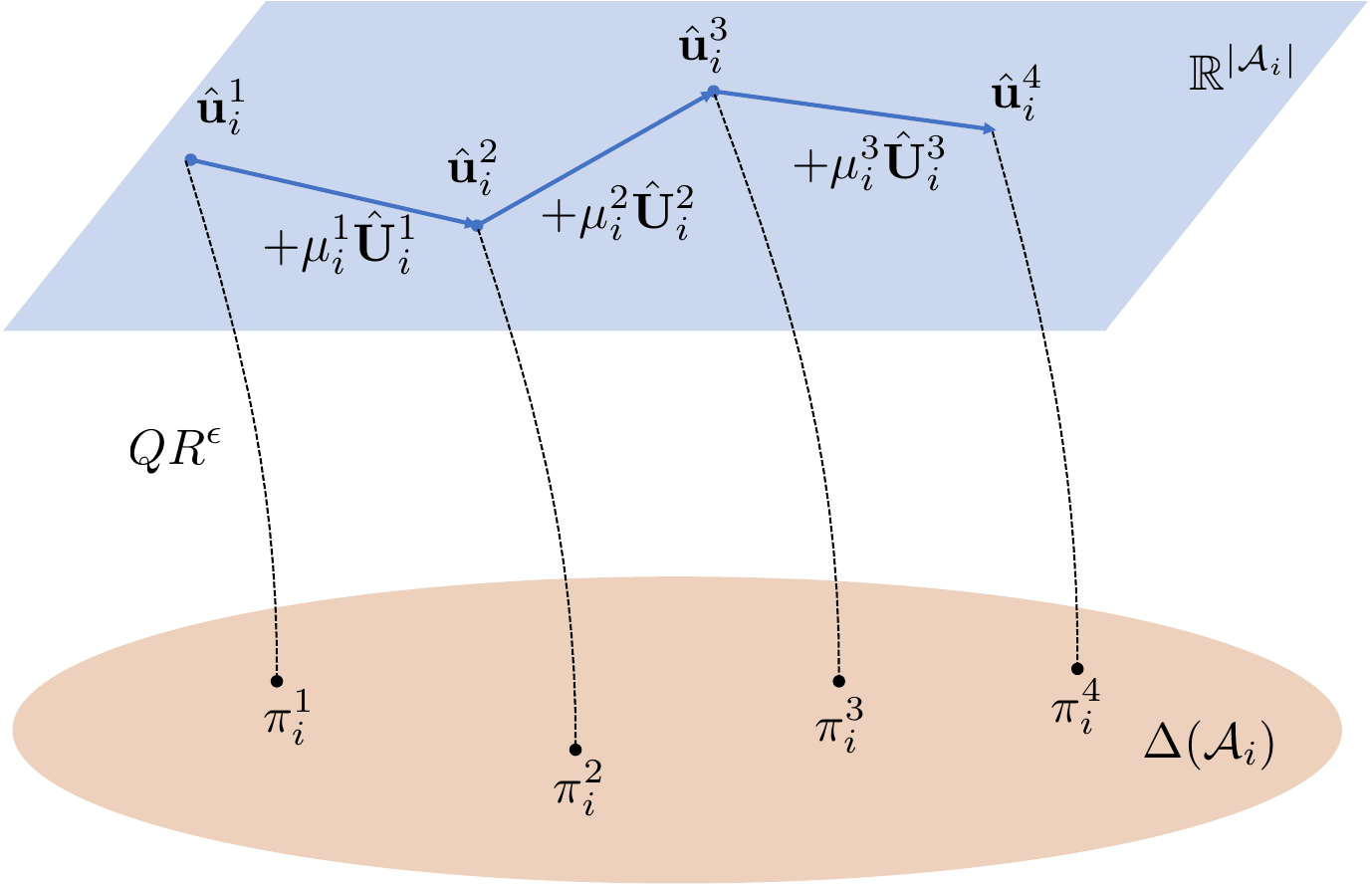}
    \caption{Schematic representation of dual averaging \eqref{eq:da}. There is no explicit dynamics in the primal space $\Delta(\mathcal{A}_i)$. Instead, the dual variables $\hat{\mathbf{U}}_i^k$ are first aggregated within the dual space $\R^{|\mathcal{A}|_i}$, and then are ``mirrored'' back to the primal space via the mirror mapping $QR^\epsilon$.   }
    \label{fig:da}
\end{figure}

In finite games, the utility function is not differentiable with respect to the action, since actions sets are finite. In order to leverage the gradient play, we consider the mixed extension of the finite games. Consider the expected utility $u_i(\pi_i,\pi_{-i})=\left\langle \pi_i, \mathbf{u}_i(\pi_{-i})\right\rangle$, then the gradient of the expected utility with respect to player $i$'s strategy $\pi_i$ is given by $\mathbf{u}_i(\pi_{-i})$. Naturally, we can apply the mirror descent scheme \eqref{eq:omd} to this mixed extension without difficulty. Furthermore, if the gradient is not directly available, for example, learning under the noisy feedback, we rely on the unbiased estimator of $\mathbf{u}_i(\pi_{-i}^k)$, $\hat{\mathbf{U}}_i^k$, which can be viewed as an estimator of the payoff gradient $D_i$ in \eqref{eq:omd}. It can be easily seen that the mirror descent scheme for this induced continuous game reduces to the exploratory reinforcement learning in \eqref{eq:da}. Consequently, the learning scheme \eqref{eq:da} is called dual averaging: the dual variables, the gradients $\hat{\mathbf{U}}_i^k$, are aggregated first within the dual space and then are ``mirrored'' back to the primal space by the mirror mapping \cite{nesterov09da}. A schematic representation of dual averaging is provided in \Cref{fig:da}.

\subsection{Convergence of Learning in Games}\label{sec:conver}
This subsection examines the asymptotic behavior of learning algorithms introduced above, with the focus on the convergence results of the introduced learning algorithms. Due to the close connection between gradient play in continuous games and reinforcement learning in finite games, we limit our scope to reinforcement learning algorithms in finite games, while pointing the reader to \cite{PM18cont,ratliff20grad_nash,bravo18bandit_learning,hofbauer06br_zero,hofbauer02sbr} for the treatment in continuous games. The discussion in this subsection is primarily based on stochastic approximation theory and Lyapunov stability theory \cite{benaim05SADI,Perkins12asy_SA}, and a generic procedure of applying such analytical tools consists of three steps: 1) develop the mean-field continuous-time dynamics using stochastic approximation theory; 2) study the continuous-time learning dynamics using ODE methods, relating its Lyapunov stability to Nash equilibria of the underlying game; 3) derive the convergence results of  discrete-time algorithms using asymptotic convergence of corresponding continuous-time dynamics. Since the third step is direct corollary of the results of the first and second steps, we articulate the first two steps in analyzing the asymptotic behaviors of reinforcement learning in the sequel. We refer the reader to \Cref{side:sa} and references therein for details on the relation between discrete-time trajectory and its continuous counterpart.
\subsubsection{Learning Dynamics and Stochastic Approximation}\label{sec:con_dyna}
With proper $F^k_i$ and $G^k_i$, learning algorithms allow the players to reach the Nash equilibrium of the game in the limit. Hence, the problem boils down to analyzing the limiting behavior of the discrete-time systems \eqref{eq:br_noisy},\eqref{eq:da},\eqref{eq:sbr}, that is, whether its global attractor comprises equilibria. Direct investigations into such learning dynamics are challenging, as stochasticity enters the updating rules. For example, the action at time $k$, $a^k_i$ is sampled from the strategy $\pi^k_i$, and the payoff feedback $U^k_i$ also incurs randomness. 

Thanks to the celebrated stochastic approximation theory, we can turn to the continuous counterpart of the discrete-time dynamics: an ordinary differential equation (ODE), whose trajectory enjoys the same asymptotic property. From a technical standpoint, the continuous-time dynamics often produce a more comprehensible picture for analysis with fruitful tools available at our disposal. One of the most powerful tools is Lyapunov stability theory.  Besides, such a continuous-time framework also allows us to connect learning theory with the extensive literature on game dynamics in biology and evolutionary theory \cite{fuden_learning}, where the time interval between two repetitions of the game is infinitesimally small.   

Recall that reinforcement learning adopts two coupled discrete-time dynamical systems: one for the score function \eqref{eq:utility} and the other one for the choice mapping \eqref{eq:strategy}.
\begin{align*}
    \hat{\mathbf{u}}_i^{k+1} &= (1-\mu_i^k)\hat{\mathbf{u}}_i^k +\mu_i^k G^k_i(\pi^k_i, \hat{\mathbf{u}}_i^k, U_i^k, a_i^k),\\
    \pi_i^{k+1}&=(1-\lambda_i^k)\pi_i^k+\lambda_i^k F_i^k(\pi_i^k,\hat{\mathbf{u}}_i^{k+1}, U_i^k, a_i^k).
\end{align*}
In the following, the continuous-time dynamics associated with \eqref{eq:utility} and \eqref{eq:strategy} is obtained via stochastic approximation, which paves the way for the ODE-based convergence analysis. We begin with a generic description of the learning dynamics under reinforcement learning, and then we specify the learning dynamics corresponding to \eqref{eq:br_noisy}\eqref{eq:da}\eqref{eq:sbr}.  For more details regarding stochastic approximation, we refer the reader to \Cref{side:sa} and the references therein.

For the sake of simplicity in exposition, we assume that learning policies in \eqref{eq:utility} and \eqref{eq:strategy} are time-invariant, denoted by $F_i$ and $G_i$, respectively. When the learning policies are time-variant, stochastic approximation theory still applies, and we refer the reader to \cite{zhu12heter} for more details. Let the mean-field components of \eqref{eq:utility} and \eqref{eq:strategy} be denoted by $f_i(\pi_i^k,\hat{\mathbf{u}}_i^{k+1})=\mathbb{E}[F_i(\pi_i^k,\hat{\mathbf{u}}_i^{k+1}, U_i^k, a_i^k)|\mathcal{F}^{k-1}]$ and $g_i(\pi_i^k,\hat{\mathbf{u}}_i^k)=\mathbb{E}[G_i(\pi_i^k,\hat{\mathbf{u}}_i^k, U_i^k, a_i^k)|\mathcal{F}^{k-1}]$, respectively. We can then write down the following coupled differential equations
\begin{align*}
\begin{aligned}
&\frac{d \hat{\mathbf{u}}_i(t)}{dt} = g_i(\pi_i(t), \hat{\mathbf{u}}_i(t)),\\
	&\frac{d \pi_i(t)}{d t}=  f_i(\pi_i(t), \hat{\mathbf{u}}_i(t)),
\end{aligned}
\end{align*}
which are closely related to \eqref{eq:utility} and \eqref{eq:strategy}. By stochastic approximation theory (see \Cref{side:sa}), the linear interpolations of the sequences $\{\pi_i^k\}$ and $\{\hat{\mathbf{u}}_i^{k}\}$ are the perturbed solutions to the differential equations above, which are arbitrarily close to the true solution as time goes to infinity. In other words, the convergence results of \eqref{eq:utility} and \eqref{eq:strategy} can be obtained by studying the limiting behavior of the associated differential equations. 

Following the same argument, the learning dynamics of the best response \eqref{eq:br_noisy} can be written as
\begin{align}
\begin{aligned}\label{eq:br_c}
	&\frac{d \hat{\mathbf{u}}_i(t)}{dt}=\mathbf{u}_i(\pi_{-i}(t))-\hat{\mathbf{u}}_i(t),\\
	&\frac{d\pi_i(t)}{dt}\in BR_i(\hat{\mathbf{u}}_i(t))-\pi_i(t).
\end{aligned}\tag{\text{BR-c}}
\end{align}
If the best response dynamics is adopted by every player, we can consider the continuous-time dynamics of the strategy profile of all players $\pi(t)=[\pi_1(t),\pi_2(t),\ldots,\pi_N(t)]$ under best response. Denote the joint utility vector by $\mathbf{u}(\pi(t)):=[\mathbf{u}_1(\pi_{-1}(t)),\mathbf{u}_2(\pi_{-2}(t)),\ldots, \mathbf{u}_N(\pi_{-N}(t))]$, and similarly, the joint estimated utility vector by $\hat{\mathbf{u}}(t):=[\hat{\mathbf{u}}_1(t),\hat{\mathbf{u}}_2(t),\ldots,\hat{\mathbf{u}}_N(t)]$. Then, for the strategy profile $\pi(t)$, the continuous-time learning dynamics under the best response algorithm is 
\begin{align}
    &\frac{d \hat{\mathbf{u}}(t)}{dt}=\mathbf{u}(\pi(t))-\hat{\mathbf{u}}(t),\label{eq:moving_dyna}\\
	&\frac{d\pi(t)}{dt}\in BR(\hat{\mathbf{u}}(t))-\pi(t).\label{eq:br_dyna}
\end{align}
 From its associated learning dynamics, we can see that the best response algorithm \eqref{eq:br_noisy} or equivalently its continuous-time mean-field dynamics \eqref{eq:br_c} is in fact an actor-critic learning \cite{Sutton:2018wc}, where the approximation $\hat{\mathbf{u}}(t)$  given by \eqref{eq:moving_dyna} serves as the actor,  evaluating the performance of the current strategy profile, while the strategy update  \eqref{eq:br_dyna} is the critic that improves the strategy. 
 
 As observed in the literature \cite{Sutton:2018wc}, {the performance of the actor-critic learning relies on the quality of evaluation from the actor.} One approach to obtain a satisfying actor in learning is to leverage the two-timescale idea \cite{Borkar:2009ts}, according to which $\eqref{eq:moving_dyna}$ should operate at a faster timescale than \eqref{eq:br_dyna}. Intuitively speaking, in order to obtain a $\hat{\mathbf{u}}(t)$ that can approximately evaluate the current strategy profile $\pi(t)$, the player must wait until $\hat{\mathbf{u}}(t)$ nearly converges before it updates the strategy using \eqref{eq:br_dyna}. To analyze the convergence of the two-timescale dynamics, one can study its equivalent single-timescale dynamics. Since the actor \eqref{eq:moving_dyna} runs at a faster timescale, the system \eqref{eq:moving_dyna} and \eqref{eq:br_dyna} can be ``decoupled'' in the following way:  {by fixing $\pi(t)=\pi$, the faster timescale update \eqref{eq:moving_dyna} converges to $\mathbf{u}(\pi)$, where $\pi$ is viewed as a parameter}, Then, {after the convergence of the fast dynamics to an equilibrium $\mathbf{u}(\pi)$},  the slow dynamics \eqref{eq:br_dyna} is set into motion, where $\hat{\mathbf{u}}(t)$ is replaced by its equilibrium point $\mathbf{u}(\pi(t))$ and the resulting learning dynamics is 
\begin{align}\label{eq:br_equiva}
    \frac{d\pi(t)}{dt}\in BR(\pi(t))-\pi(t).
\end{align}  
As we illustrate in  \Cref{side:sa}, the coupled dynamics \eqref{eq:moving_dyna}\eqref{eq:br_dyna} and the single-timescale \eqref{eq:br_equiva} share similar asymptotic behaviors. Hence, we can focus on the much simplified one \eqref{eq:br_equiva} for the derivation of the convergence results. For more details about the two-timescale learning and the derivation of the equivalent dynamics, we refer the reader to \Cref{side:sa} and references therein.

Applying the same argument to the smoothed best response \eqref{eq:sbr}, we obtain 
\begin{align}\label{eq:sbr_c}
    \begin{aligned}
    &\frac{d \hat{\mathbf{u}}_i(t)}{dt}=\mathbf{u}_i(\pi_{-i}(t))-\hat{\mathbf{u}}_i(t),\\
    	&\frac{d\pi_i(t)}{dt}=QR^\epsilon(\hat{\mathbf{u}}_i(t))-\pi_i(t),
\end{aligned}\tag{\text{SBR-c}}
\end{align} 
and its equivalent dynamics regarding the joint strategy profile is 
\begin{align}\label{eq:sbr_equiva}
    \frac{d\pi(t)}{dt}= QR^\epsilon(\mathbf{u}(\pi(t)))-\pi(t).
\end{align}

Different from the best response \eqref{eq:br_noisy} and the smoothed best response \eqref{eq:sbr}, dual averaging \eqref{eq:da} does not belong to the class of actor-critic methods. To see this, let us write down its continuous-time dynamics 
\begin{align}\label{eq:da_con}
\begin{aligned}
	&\frac{d {\hat{\mathbf{u}}}_{i}(t)}{dt}=\mathbf{u}_i(\pi_{-i}(t)),\\
	& \pi_i(t)=QR^\epsilon(\hat{{\mathbf{u}}}_i(t)).
\end{aligned} \tag{\text{DA-c}}
\end{align}
Similar to the previous argument, the learning dynamics for the strategy profile is 
\begin{align}\label{eq:cda}
\begin{aligned}
    \frac{d \hat{\mathbf{u}}(t)}{dt}&= \mathbf{u}(\pi(t)),\nonumber\\
	\pi(t)&= QR^\epsilon(\hat{\mathbf{u}}(t)),\nonumber
\end{aligned}\tag{\text{DA}}
\end{align}
where the dynamics regarding $\hat{\mathbf{u}}(t)$ does not produce an approximation of $\mathbf{u}(\pi(t))$. Instead, it gives the cumulative payoff: $\hat{\mathbf{u}}(t)=\int_0^t \mathbf{u}(\pi(\tau)) d\tau+\hat{\mathbf{u}}(0)$. It is straightforward to see that as there is only one differential equation in  \eqref{eq:cda},  the resulting autonomous dynamical system is only related to $\hat{\mathbf{u}}(t)$. Hence, there is no additional dynamics regarding the strategy update, which makes \eqref{eq:cda} fundamentally different  from \eqref{eq:br_c} and \eqref{eq:sbr_c}.

\subsubsection{Nash Equilibrium and Lyapunov Stability}\label{sec:nash_lyap}

Since the various learning algorithms belong to different classes,  the discussion regarding the convergence results of the introduced learning dynamics are organized in the following way. We begin with dual averaging \eqref{eq:cda}, a type of gradient-based dynamics, and then proceed  to the best response dynamics \eqref{eq:br_c} and the smoothed best response \eqref{eq:sbr_c}.

\paragraph{Dual Averaging}  Consider the learning dynamics of the joint strategy profile and the estimated utility vector under dual averaging
\begin{align}
\begin{aligned}
	\frac{d \hat{\mathbf{u}}(t)}{dt}&= \mathbf{u}(\pi(t)),\\
	\pi(t)&= QR^\epsilon(\hat{\mathbf{u}}(t)).
\end{aligned}\tag{\text{DA}}
\end{align}   
This compact form implies  that \eqref{eq:cda} is an autonomous system evolving in the dual space. Here, similar to the discussion in \Cref{sec:gp}, we adopt the terminology in \cite{nesterov09da,nesterov04book}, where the gradient $\mathbf{u}(\pi(t))$ is the dual variable and the corresponding space is termed the dual space.  As shown in  \cite{PM16lyapnov}, \eqref{eq:cda} is a well-posed dynamical system in the dual space in that it admits a unique global solution for every initial $\hat{\mathbf{u}}(0)$. Furthermore, it can be shown that the dynamics of $\pi(t)$ on the game's strategy space induced by \eqref{eq:cda} under steep regularizers is also well-posed \cite{PM16lyapnov,PM18rieman}. However, the well-posedness of the induced dynamics under generic regularizers remains unclear \cite{PM16lyapnov}. The reason lies in the fact that under steep regularizers, such as the entropy regularizer,  the projected dynamics regarding $\pi(t)$ evolves within the interior of the simplex, and the resulting ODE is also well posed in the primal space, which need not hold for nonsteep regularizers. For more generic choices of $QR$ and related stability analysis, we refer the reader to \cite{PM16lyapnov}. 

Even though studying the stability of the induced dynamics in the primal space may not be viable due to the well-posedness issue,  the {asymptotic} behavior of $\pi(t)$ can be characterized by investigating its dual $\hat{\mathbf{u}}(t)$. Toward that end, we call $\pi(t)=QR^\epsilon(\hat{\mathbf{u}}(t))$ the induced orbit of \eqref{eq:cda} or simply orbit, and we introduce the following notions regarding the stability and stationarity of $\pi(t)$, which is adapted from \cite{PM16lyapnov}.  
\begin{definition}
Denote by $\operatorname{im}(QR^\epsilon)$ the image of $QR^\epsilon$. For $\pi(t)=QR(\mathbf{u}(t))$, an orbit of \eqref{eq:cda}, we say that a fixed $\pi^*\in\prod_{i\in \mathcal{N}} \Delta(\mathcal{A}_i) $ is 
	\begin{enumerate}
		\item \textit{stationary}, if $\pi(t)=\pi^*\in \operatorname{im}(QR^\epsilon)$ for all $t\geq 0$, whenever $\pi(0)=\pi^*$;
		\item \textit{Lyapunov stable}, if for every neighborhood $U$ of $\pi^*$, there exists a neighborhood $U'$ of $\pi^*$ such that $\pi(t)\in U$ for all $t\geq 0$ whenever $\pi_0\in U'\cap \operatorname{im}(QR^\epsilon)$;
		\item \textit{attracting}, if there exists a neighborhood $U$ such that $\pi(t)\rightarrow \pi^*$ as $t\rightarrow\infty$ whenever $\pi_0\in U\cap\operatorname{im}(QR^\epsilon)$;
		\item \textit{globally attracting}, if $\pi^*$ is attracting with the attracting basin being the entire image $\operatorname{im}(QR^\epsilon)$;
		\item \textit{asymptotically stable}, if $\pi^*$ is both attracting and Lyapunov stable; 
		\item \textit{globally asymptotically stable}, if $\pi^*$ is both globally attracting and Lyapunov stable.
	\end{enumerate}
\end{definition}

Similar to the Folk Theorem of evolutionary game theory \cite{hofbauer03evolu_dyna},  there is an equivalence between the stationary points of \eqref{eq:cda} and the Nash equilibria \cite{PM16lyapnov, hofbauer03evolu_dyna}: any stationary point is a Nash equilibrium and conversely, every Nash equilibrium that is within the image of the mirror map \eqref{eq:mirror} is a stationary point. In addition to the relation between Nash equilibrium and the stationary point, another important question is the following:
\begin{center}
    \textit{Are Nash equilibria of the underlying game (globally) asymptotically stable under  \eqref{eq:cda}?}
\end{center}
To answer this question, we shall revisit the variational characterization of Nash equilibrium, which bridges the equilibrium concepts associated with two different mathematical models: games and dynamical systems. Recall that the Nash equilibrium is equivalent to the solution of the variational inequality 
\begin{align}\label{eq:svi}
    \langle \mathbf{u}(\pi^*), \pi-\pi^*\rangle \leq 0, \quad \text{for all } \pi\in \prod_{i\in \mathcal{N}} \Delta(\mathcal{A}_i). \tag{SVI}
\end{align}
Since the utility function $u_i(\pi_i,\pi_{-i})$ is linear in $\pi_i$, the Stampacchia-type variational inequality \eqref{eq:svi} is equivalent to the following Minty-type variational inequality
\begin{align}\label{eq:mvi}
    \langle \mathbf{u}(\pi), \pi-\pi^*\rangle \leq 0, \quad \text{for all } \pi\in \prod_{i\in \mathcal{N}} \Delta(\mathcal{A}_i), \tag{MVI}
\end{align}
which implies that the Nash equilibrium $\pi^*$ is the solution to \eqref{eq:mvi} \cite{pappallardo02nash_vi}. Then, to answer the question of interest, it suffices to investigate whether the solution to \eqref{eq:mvi} is attracting under \eqref{eq:cda}. As discussed in \cite{ratliff20grad_nash}, the answer is negative: not every Nash equilibrium of $N$-player general-sum game is attracting. To ensure the convergence of \eqref{eq:cda}, an additional condition has to be imposed on \eqref{eq:mvi}. 
\begin{definition}[Variational Stability \cite{PM16lyapnov}]
$\pi^*$ is said to be variationally stable if there exists a neighborhood $U$ of $\pi^*$ such that 
\begin{align}\label{eq:vs}
     \langle \mathbf{u}(\pi), \pi-\pi^*\rangle \leq 0, \quad \text{for all } \pi\in U, \tag{VS}
\end{align}
where equality holds if and only if $\pi^*=\pi$. In particular, if $U=\prod_{i\in \mathcal{N}}\Delta(\mathcal{A}_i)$, $\pi^*$ is said to be globally variationally stable. 
\end{definition}
The definition of variational stability (VS) can be extended to sets \cite{PM16lyapnov}. Let a subset $\Pi^*\subset\prod_{i\in \mathcal{N}}\Delta(\mathcal{A}_i)$ be closed and nonempty. $\Pi^*$ is said to be variationally stable if there exists a neighborhood $U$ of $\Pi^*$ such that 
\begin{align}
     \langle \mathbf{u}(\pi), \pi-\pi^*\rangle \leq 0, \quad \text{for all } \pi\in U, \pi^*\in \Pi^*, 
\end{align}
where equality holds for a given $\pi^*\in \Pi^*$ if and only if $\pi\in \Pi^*$. 

The notion ``variational stability'' is proposed in \cite{PM16lyapnov} as a relaxation of the monotonicity condition of the pseudo-gradient mapping of the game, e.g., $\mathbf{u}(\pi)$ in the mixed extension of finite games, or $D(\mathbf{a})$ in continuous games. Variational stability alludes to the seminal notion of evolutionary stability introduced in \cite{maynard73logic}, and the introduced definition is in a similar spirit to the variational characterization of evolutionarily stable state studied in \cite{hofbauer03evolu_dyna}. An equivalent notion is developed in the line of works on gradient-based learning \cite{ratliff20grad_nash}, named \textit{locally asymptotically stable Nash equilibria (LASNE)}, and as its name suggests, Nash equilibria satisfying the variational stability (VS) are asymptotically stable under gradient-based dynamics. Likewise, Nash equilibria satisfying global variational stability are globally asymptotically stable (GASNE). We refer the reader to \cite{ratliff20grad_nash} and references therein for more details about this characterization of Nash equilibria.

What has been presented above provides a generic criterion for examining the convergence of gradient-based dynamics \eqref{eq:cda}, and in the following, based on the notion of variational stability, we discuss some concrete cases, where the learning dynamics converges either locally or globally to Nash equilibria. As shown in \cite{PM18cont}, for any finite games, every strict Nash equilibrium satisfies \eqref{eq:vs} and hence is a LASNE. Therefore, every strict Nash equilibrium in finite games is locally attracting. On the other hand, to ensure global convergence, the underlying Nash equilibrium has to be GASNE or equivalently satisfy the global variational stability. For finite games, the existence of a potential implies monotonicity, which further implies the existence of globally variationally stable Nash equilibiria \cite{PM18cont}. Hence, for potential games \cite{nips2017_7216,PM16lyapnov} and monotone games \cite{pavel19passivity,PM18cont}, regardless of the initial points, the orbit of \eqref{eq:cda} always converges to the set of Nash equilibria. We summarize our discussions in the following, where 1) and 2) are direct extensions of the folk theorem of evolutionary dynamics \cite{hofbauer03evolu_dyna}, while 3)-5) are corollaries of variational characterization of Nash equilibria in  \cite{PM16lyapnov} and \cite{ratliff20grad_nash}.

For every finite game, we have the following characterization of Nash equilibrium using the language of Lyapunov stability \cite{PM16lyapnov,PM18rieman}. For a fixed $\pi^*\in \prod_{i\in \mathcal{N}} \Delta(\mathcal{A}_i)$, 
	\begin{enumerate}
		\item if $\pi^*$ is stationary, it is a Nash equilibrium;
		\item if $\pi^*$ is Lyapunov stable, then $\pi^*$ is a Nash equilibrium;
		\item if $\pi^*$ is a Nash equilibrium and it falls within the image of the mirror map, then it is stationary;
		\item if $\pi^*$ is a strict Nash equilibrium, it is asymptotically stable;
		\item if $\pi^*$ is a Nash equilibrium of a potential game or a monotone game, it is globally asymptotically stable.
	\end{enumerate}

\paragraph{Best response dynamics} The analysis of the best response dynamics \eqref{eq:br_equiva} is more involved than that of dual averaging \eqref{eq:cda}. The theoretical challenge is mainly due to the discontinuous, set-valued nature of the best response mapping \eqref{eq:br_mapping}. In general, as a differential inclusion,  \eqref{eq:br_equiva} typically admits non-unique solutions through every initial point \cite{benaim05SADI}. Early works have established the convergence results on \eqref{eq:br_equiva} for games with special structures: best response dynamics converges to Nash equilibrium in zero-sum games, where the Nash equilibrium is essentially a saddle point \cite{harris98con_fp,hofbauer06br_zero,barron10brd},  in two-player strictly supermodular games \cite{sandholm10evolu_game} and in finite potential games \cite{harris98con_fp,benaim05SADI}. However, we note that these research works, even though most of them still rely on the Lyapunov argument \cite{benaim05SADI,harris98con_fp,hofbauer06br_zero,barron10brd}, do not directly reveal any generic relation between Lyapunov stability and Nash equilibrium in general multi-player non-zero sum games, and they are more or less on an ad hoc basis.  

Recent endeavors on the study of the best response dynamics have helped shed some light on the asymptotic behavior of best response dynamics by relating the best response vector field $BR(\pi)-\pi$ to the gradient field $\mathbf{u}(\pi)$, which renders the best response dynamics in some potential games \cite{swenson18br_potential,swenson20regular} as an approximation of the gradient-based dynamical system \cite{swenson18br_potential}. For the finite potential games considered in \cite{swenson18br_potential}, additional regularity conditions are imposed, which are closely related to the notion of variational stability introduced above. Therefore, the variational characterization of Nash equilibrium and varitional stability becomes relevant under the best response dynamics. Following this line of reasoning, it is shown in \cite{swenson18br_potential}, in regular potential games, that the best response dynamics is well-posed for almost every initial condition, and converges to the set of Nash equilibria. 

\paragraph{Smoothed Best Response} As we can see from the explicit expression, smoothed best response dynamics \eqref{eq:sbr_equiva} only differs from the best response dynamics \eqref{eq:br_equiva} in the operator $QR^\epsilon(\cdot)$, which serves as a perturbed best response \cite{benaim12perturb_br}, and the perturbation is determined by $\epsilon$ \cite{horbauer09time}. Hence, if $\epsilon$ tends to zero, it is straightforward to see that the smoothed best response \eqref{eq:sbr_equiva} will enjoy the same asymptotic property as the best response \eqref{eq:br_equiva}, which implies that identical results should also be achievable for smoothed best response with vanishing exploration. This intuition has been verified in \cite{hofbauer02sbr,leslie06sbr}, where smoothed best response \eqref{eq:sbr_equiva} is shown to converge in zero-sum games, potential games and supermodular games. 

On the other hand, with a constant $\epsilon$, it is not realistic to expect the smoothed best response, essentially a fixed point iteration, to always converge to exact Nash equilibrium.  Hence, a new equilibrium concept has been introduced in the literature, which is termed perturbed Nash equilibrium in \cite{harsanyi,hofbauer05pert_NE} or Nash distribution in \cite{leslie03sbr_multi,leslie05sbr_q}. The new equilibrium is defined as the fixed point of the smoothed best response. We do not carry out detailed discussion on that in this paper, since the convergence analysis still rests on the standard Lyapunov argument, and the epistemic justification of such equilibrium \cite{harris98con_fp,fuden_learning} is beyond the scope of this paper. We refer the reader to \cite{fuden_learning,hofbauer02sbr,hofbauer05pert_NE,leslie05sbr_q} for a rigorous treatment of the smoothed best response.

\subsection{Beyond Stochastic Approximation}
 In addition to stochastic approximation and related ODE methods, another class of widely applied learning algorithms is built upon Markov Chain theory \cite{young93perturbMC}, which is termed learning by trial and error (LTE) \cite{young09lte}.    Even though the name of the proposed learning suggests its similarity to reinforcement learning, the learning process is quite different in the sense that there are no explicit score functions or choice mappings in the proposed method. In LTE, there are two basic rules: 1) players occasionally experiment with alternative strategies, keeping the new strategy if, and only if, it leads to a strict increase in payoff; 2) if the player experiences a payoff decrease due to a strategy change by someone else, it starts a random search for a new strategy, eventually settling on one with a probability that increases monotonically with its realized payoff. In words, the ``error'' part relies on the realized payoff, and no advanced device is needed, such as score functions like Q-functions or estimated utilities, while the ``trial'' part is a random search procedure implemented according to the two basic rules. A novel feature of the process is that different search procedures are triggered by different psychological states or \textit{moods}, where mood changes are induced by the relationship between a player’s realized payoffs and his current payoff expectations. To be specific, there are four  moods: \textit{Content}(\textit{C}), \textit{Hopeful}(\textit{H}), \textit{Watchful}(\textit{W}) and \textit{Discontent}(\textit{D}), and different moods lead to different random search procedures. Briefly, players will explore new strategies with high probabilities when in W and D, while sticking to the current one, with high probabilities, if the mood is C or H. Details can be found in the original paper, and a concise summary is provided in \cite{gaveau20perf_tel}. 

This mood-based trial and error is different from reinforcement learning introduced in the previous subsection, where the exploration is not determined eplicitly by the score function and the choice mapping. Hence, LTE does not fit the stochastic approximation framework introduced above, and instead, the associated convergence proof relies on perturbed Markov Chain theory \cite{young93perturbMC,marden09safe_exp}. It is shown in \cite{young09lte} that in a two-player finite game, if there at least exists a pure Nash equilibrium, then LTE guarantees that pure Nash equilibrium is played at least $1-\epsilon$ of the time, where $\epsilon$ is the probability of exploring new strategies. For an $N$-player finite game, if the game is \textit{interdependent} \cite{young09lte} and there at least exists one pure Nash equilibrium, the same  theoretical guarantee for the two-player case also holds. It is not surprising that LTE does not achieve convergence in conventional ways, that is, almost sure convergence and convergence in the mean, since players will always explore new strategies with positive probability at least $\epsilon$. The proposed learning method and its variants have also been applied to learning efficient equilibrium \cite{young12efficientNE} (Pareto dominant, maximizing social welfare), learning efficient correlated equilibria  \cite{marden17effi_ce}, achieving the Pareto optimality \cite{marden14pareto} and other related works in engineering applications, especially in cognitive radio problems \cite{marden14review_young}. 

The idea of trial and error in LTE leads to many important variants, such as sample experimentation dynamics in \cite{marden09safe_exp} and optimal dynamical learning \cite{marden14pareto,gaveau20perf_tel}, which also rely on perturbed Markov processes for equilibrium seeking.  Even though the convergence results of these algorithms all rest on Markov Chain (MC) theory \cite{young93perturbMC}, the analysis of their performance remains unclear, due to the computation complexity of the inherent MC generated by these algorithms. To circumvent the dimensionality issue regarding the number of states in the original MC, an approximation-based dimension reduction method is proposed in \cite{gaveau20perf_tel}, which allows numerical convergence analysis for LTE and its variants based on Monte Carlo simulations. Besides, we also note that a much simplified trial-and-error algorithm has been theoretically analyzed in \cite{minghui19}, where the optimal exploration rate is identified and the associated convergence rate is discussed. It is not unrealistic to expect a similar argument may apply to LTE and its variants, but the technical challenges regarding the dimensionality should not be downplayed.
\subsection{Resurgence of Learning in Games}
With machine learning (ML) algorithms being increasingly deployed in real-world applications, there has been a resurgence in research endeavors on multi-agent learning and learning in games \cite{kaiqing_overview}. In addition to the line of research driven by evolutionary dynamics dating back to 1950s \cite{sandholm10evolu_game,hofbauer03evolu_dyna}, the current wave of learning theory development is mainly driven by a desire to better understand and improve the performance of ML algorithms in a competitive environment. In general, there are two possible roles that game theoretic methods can play in ML study: 1) Game-theoretic methods is an add-on for improving the performance of ML algorithms. 2) Certain ML problems manifest the game features, which calls for game-theoretic tools.  For supervised learning, the recent interest in adversarial learning techniques serves as an example to show how game-theoretic models and learning methods can be used to robustify machine learning \cite{zhou18sur_adgame,kaiqing_robust}, where potential attacks or disturbance are viewed as strategic moves of an opponent. On the other hand, there are problems in unsupervised learning where game-theoretic models are no longer tools for solving the problem but the problem itself. Generative Adversarial Networks (GAN) \cite{gan_goodfellow}, is an approach to generative modeling using deep learning methods, involving automatically discovering and learning the patterns of input data in such a way that newly generated examples output by the generative model (generator) cannot be distinguished from the input. In game-theoretic language, the training process of GAN is essentially a learning process in a zero-sum game between the generator and the discriminator, where the generator tries to generate new samples that plausibly could have been drawn from the original dataset, while the discriminator tries to pick those fake ones produced by the generator. We do not intend to provide a comprehensive survey for these machine learning applications, instead we refer the reader to \cite{kaiqing_overview,zhou18sur_adgame}.    

Despite different contexts under which the learning theory is studied, recent research efforts mainly revolve around the following three aspects:
\begin{enumerate}
    \item learning dynamics in general multi-player repeated games;
    \item learning dynamics in repeated games with acceleration design;
    \item learning dynamics in dynamic games in a decentralized manner.
\end{enumerate}
The first research direction is a natural follow-up to the study of evolutionary dynamics \cite{sandholm10evolu_game,hofbauer03evolu_dyna}, which aims to bring learning in games to a broad range of ML applications, since in ML, the game structure is specified by the underlying data and may not enjoy any desired properties. Recall that convergence results and asymptotic behaviors regarding the three dynamics \eqref{eq:br_c}\eqref{eq:sbr_c}\eqref{eq:da_con} are discussed with the assumption that the underlying game acquires special structures, such as potential games, supermodular games and zero-sum games. However, for games with fewer assumptions on the utility function, there is still a lack of understanding of the dynamics and the limiting behavior of learning algorithms. One of the central questions of this direction is \textit{what the relations between Nash equilibria and stationary points as well as attracting sets under the learning dynamics are.} Recent attempts try to answer this question from a variational perspective \cite{wibisono16var}, and provide various characterizations of Nash equilibria with desired properties under gradient-based dynamics \cite{mazumdar19localNE,ratliff20grad_nash,PM18rieman}. Furthermore, considering its applications in ML problems, learning algorithms in stochastic settings are of great significance in recent studies, and we refer the reader to \cite{chijin19nonconvex,ratliff20grad_nash} for more details as well as to \cite{lei_csm} for an introduction to stochastic Nash equilibrium seeking. 

The second research direction, which attracts attention from the ML community, the optimization community as well as the control community, is directly related to the design of ML algorithms. The goal is to develop acceleration techniques that improve the performance of learning algorithms. Based on the understanding of first-order gradient-based dynamics games such as \eqref{eq:pgd}\eqref{eq:lgd}, recent research efforts have focused on high-order gradient methods, which can be dated back to Nesterov's momentum idea \cite{nesterov04book}, and researchers endeavor to propose a general framework that generalize the momentum for the generation of accelerated gradient-based algorithms \cite{wibisono16var}.  On account of the close relationships among Nash equilibrium, variational problems and dynamical systems \cite{pappallardo02nash_vi}, one approach for developing acceleration is to generalize the concept momentum by formulating the equilibrium seeking as a variational (optimization) problem \cite{pappallardo02nash_vi,diakonikolas21a}, and then investigate acceleration methods within the optimization context using, for example, variational analysis \cite{wibisono16var}, extra-gradient \cite{diakonikolas21a} and differential equation \cite{su16diff_acc}. In addition to these mentioned research works, we refer the reader to \cite{pavel_csm} for a review on the optimization-based approach.   On the other hand, as depicted in \Cref{fig:feedback}, a learning process in general is a feedback system, and it is not surprising that control theory can play a part in designing the acceleration. For example, recent studies on reinforcement learning demonstrate that passivity-based control theory can be leveraged in designing high-order learning algorithms \cite{pavel17pass,pavel19passivity}, where the learning rule is treated as the control law to be designed. Another paper \cite{basar87relax_memory} promotes the use of memory in best response maps to accelerate convergence in Nash seeking, and demonstrates substantial improvements in doing so. In addition to the mentioned references, we further refer the reader to \cite{hu_csm} for a review on control-theoretic approaches on distributed Nash equilibrium seeking, and to \cite{frihauf12extreme} for the use of extreme seeking in the learning process.

The recent advance on the third research direction is in part driven by multi-agent reinforcement learning and its applications such as multi-agent robotic control \cite{Stone2000,kaminka10,inujima13}. Different from the first two directions where the learning dynamics is primarily studied in the context of repeated games, the third research direction focuses on games with dynamic information (see \Cref{sec:dynamic_game}). In this context, the appropriate learning objective, out of practical consideration \cite{littman94markov_game}, is to obtain stationary strategies that are subgame perfect \cite{Wei17MPE} (see \Cref{sec:dynamic_game} for the definition of subgame perfectness). Different from the first two where the change to payoffs resulted from a certain action completely comes from the opponents' move,  in dynamic games, the feedback each player receives not only depends on other players' moves but also the dynamic environment.  Moreover, when making decisions at each state, players have to trade off current stage payoff for estimated future payoffs while forming predictions on the opponent’s strategies. Dynamic trade-off makes the analysis of learning in stochastic games potentially challenging \cite{sayin20fp_sg}. 

Earlier works on seeking for such Markov perfect Nash equilibrium are largely based on dynamic programming \cite{bellman:1954uq,hu03nashQ}, which requires a global information feedback, a restrictive assumption in practice. Recent efforts focus on various approaches to lessen this requirement. Currently, there are mainly three lines of research regarding learning in dynamic games. The first approach is to extend learning dynamics in repeated games to dynamic games. Built upon similar ideas in best response dynamics \eqref{eq:br_noisy}, two-timescale best response dynamics for zero-sum Markov games have been considered in \cite{leslie20br,sayin20fp_sg}, meanwhile the gradient play has been investigated in linear-quadratic dynamic zero-sum games \cite{ratliff20grad_nash,bu19pg_lqg,kaiqing21derivative_free}. The key challenge in the approach, particularly in the case of Markov games, is to properly construct the score function, which balance current stage payoffs and the future payoffs, and we refer the reader to the mentioned references for more details and to \cite{kaiqing_overview} for an overview. The second approach is to extend learning methods in single-agent Markov decision process to Markov games. However, the direct extension of methods such as Q-learning \cite{Watkins:1992jx}, policy gradient \cite{Sutton:2018wc} and actor-critic \cite{konda99two} often fail to deliver desired results due to the non-stationarity issue \cite{hernandez17non_stat}.  One natural way to overcome the non-stationarity issue is to allow players to exchange information with neighbors \cite{wai18double_aver,zhang18fully}, by which enables players to jointly identify the non-stationarity created by the dynamic environment. For more details regarding this approach, we refer the reader to recent reviews \cite{kaiqing_overview,hernandez17non_stat}.  Finally, the third approach is about a unilateral viewpoint of dynamic games. Different from the first two approaches where learning processes are still investigated in a competitive environment, the third one interprets learning in Markov games as an online optimization problem \cite{kash20no_regretQ,li2020blackwell}, where players independently make decisions based on the received feedback. This approach accounts for the fully decentralized learning, where from each player's perspective, other players are considered as part of the environment. The key idea of this approach is to leverage the regret minimization technique \cite{shai_online}, which has led to many successes in solving extensive form games of incomplete information \cite{nips2007_3306}. Despite recent advances regarding the first two approaches \cite{kaiqing_overview,leslie20br,sayin20fp_sg,ratliff20grad_nash,kaiqing20model_minmax} and positive results for the last one \cite{kash20no_regretQ,li2020blackwell,hakami}, we still lack a unified framework and a through understanding regarding the learning process in general Markov games, which remains an open area for researchers from diverse communities.

\section{Game-Theoretic Learning over Networks}\label{sec:learn_net}
Learning in games is not only intellectually interesting but also practically useful. When combined with game-theoretic modeling, such learning methods, thanks to their decentralized and adaptive nature, provide a comprehensive tool kit for designing resilient, agile, and computationally efficient controls or mechanisms for diverse applications of networks. 

In this section, we demonstrate that such a combination of game-theoretic models and associated learning dynamics, referred to as game-theoretic learning, has become indispensable for modern network problems. On the one hand, these networks often admit complex topological structures and heterogeneous nodes, resulting in large-scale complex systems, making centralized controls or mechanisms either impractical or costly. By contrast, game-theoretic models treat each node in the network as a rational and self-interested player, and the heterogeneous nature is captured by players' distinct utilities and action sets as well as information available to them, leading to a bottom-up approach for designing decentralized and scalable mechanisms and controls. On the other hand, modern networked systems, such as wireless communication networks and the smart grid, operate in a dynamic or an adversarial environment, calling for learning-based mechanisms that are responsive to changes in the environment or malicious attacks from adversaries. As shown in the last section, game-theoretic learning provides a self-adaptive procedure for each player in the system, according to which players adjust their moves based on feedback from the environment, resulting in desired collective behaviors.      

Thanks to its advantages over the centralized approach,  game-theoretic learning has gained much popularity among researchers working on multi-agent systems and network applications. There have been numerous encouraging successes in many fields, ranging from wireless and IoT communication networks \cite{alpcan2002cdma,zhu2011dynamic,zhu2009game,farooq2018secure,farooq2019modeling}, the smart grid and power networks \cite{chen2016game,maharjan2013dependable,maharjan2015demand,zhu2012differential}, infrastructure systems \cite{chen2019dynamic,huang2017large,chen2016interdependent,chen2017heterogeneous}, to cybersecurity applications \cite{xu2017game,zhu2011eavesdropping,zhu2013game,zhu2013game,huang2020dynamic,zhu2018multi}. In the following, some representative works in these fields are presented. To be specific, the focus of this section is on the applications of learning methods in wireless communications, the smart grid, and distributed machine learning, while other related applications will be briefly discussed at the end of the section.

\subsection{Next-Generation Wireless Networks}
The next-generation wireless communication technologies offer an accommodating and adaptive solution that meets the requirements of a diverse range of use cases within a common network infrastructure, providing the necessary flexibility for service heterogeneity and compatibility \cite{han_niyato_saad_basar_2019}. Such architecture, as pointed out in \cite{5g_intro}, aims to meet following demands:
\begin{itemize}
    \item increased indoor and small cell/hotspot traffic, which will make up the majority of mobile traffic volume, leading to complex network structures;
    \item higher numbers of connected heterogeneous devices stemming from the Internet of
Things (IoT), which will support massive machine-to-machine (M2M) communications and applications;
    \item improved energy consumption or efficient power control for reducing carbon footprint.
\end{itemize}

\begin{figure}
    \centering
    \includegraphics[width=\textwidth]{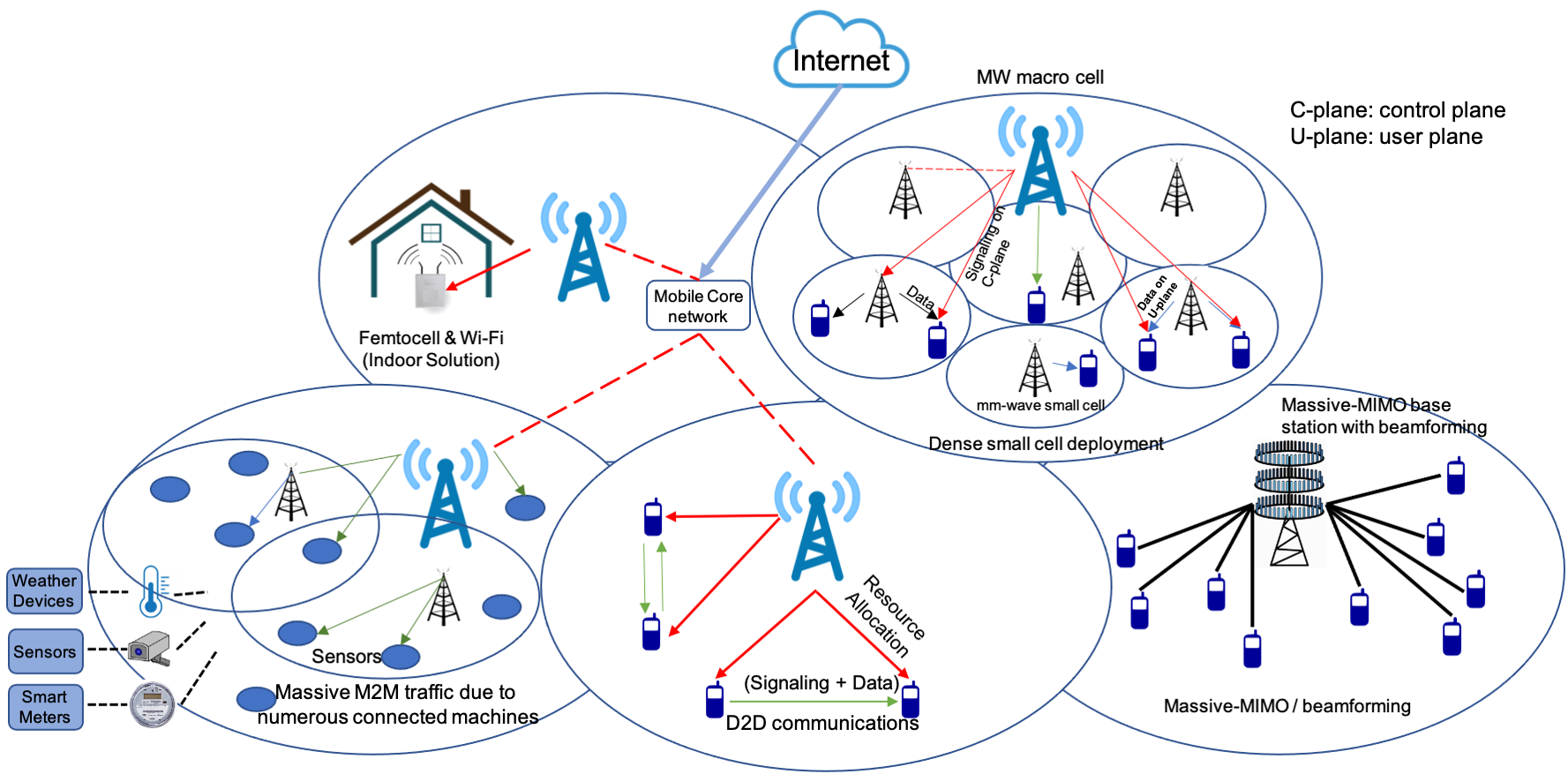}
    \caption{The next generation of communication network: macrocells (bands $<3$ GHz); small cells (millimeter-wave); femtocells and Wi-Fi (millimeter-wave); massive multiple-input, multiple-output with beamforming; and device-to-device (D2D) and machine-to-machine (M2M) communications. Solid arrows indicate wireless links, whereas the dashed arrows indicate backhaul links.}
    \label{fig:5g}
\end{figure}

From a system science perspective, these requirements impose a large-scale, time-variant, and heterogeneous network topology on modern wireless communication systems, as shown in \Cref{fig:5g}. Hence, it is impractical to manage/secure the wireless communications network centrally. Game-theoretic learning provides a scalable distributed solution with adaptive attributes to deal with this challenge. In the following, we take the dynamic secure routing mechanism as an example to illustrate how game-theoretic learning contributes to a resilient and agile communication system. 
 
Security of routing in a distributed cognitive network (CR) is a prime issue, as the routing may be compromised by unknown attacks, malicious behaviors, and unintentional misconfigurations, which makes it inherently fragile. Even with appropriate cryptographic techniques, routing in CR networks is still vulnerable to attacks in the physical layer, which can critically compromise performance and reliability. Most of the existing work focuses on the resource allocation perspective, which fails to capture the user's lack of knowledge of the attacker due to the distributed mechanism. To address these issues, \cite{zhu2011dynamic} provides a learning-based secure scheme, which allows the network to defend against unknown attacks with a minimum level of deterioration in performance.

Consider $\mathcal{G}_w:=\left(\mathcal{N}_w,\mathcal{E}_w\right)$, which is a topology graph for a multi-hop CR network, where $\mathcal{N}_w=\{n_1,n_2,...,n_N\}$ is a set of secondary users, and $\mathcal{E}_w$ is a set of links connecting these users. The system state $s$ indicates whether the primary users occupy nodes. The objective of the secondary user is to find an optimal path to its destination. In multi-hop routing, a secondary user $n_i$ starts with exploring neighboring nodes that are not occupied and then chooses a node among them to which the user routes data. The selected node initializes another exploration process for discovering the next node, and the same process is repeated until the destination is reached.

Let $\mathcal{P}_i(0,L_i):=\{(n_i,l_i),l_i\in\{0,1,2,...,L_i\}\}$ be the multi-hop path from the node $n_i$ to its destination, where $L_i$ is the total number of explorations until it reaches its destination. Suppose there are $J$ jammers in the network, the set of which is given by $\mathcal{J}:=\{1,2,...,J\}$. Let $\mathcal{R}_j$, $j\in\mathcal{J}$, be the set of nodes under the influence of jammer $j$. Denote the joint action of the jammers by $\mathbf{r}=[r_j]_{j\in\mathcal{J}}$, where $r_j\in\mathcal{R}_j$. A zero-sum game formulation is proposed in \cite{zhu2011dynamic}, where the secondary users aim to find an optimal routing path by selecting $\mathcal{P}_i(0,L_i)$, while the jammers aim to compromise the data transmission by choosing $\mathbf{r}$. The expected utility function is 
\begin{equation*}
    \mathbb{E}_s[u_i(s,\mathcal{P}_i(0,L_i),\mathbf{r})]=-\mathbb{E}_s\left[\sum_{l_i=1}^{L_i}\left(\ln q_{(n_i,l_i-1)}^{(n_i,l_i)}+\lambda \tau_{(n_i,l_i-1)}^{(n_i,l_i)}\right)\right],
\end{equation*}
where  $q_{(n_i,l_i-1)}^{(n_i,l_i)}$ is the probability of successful transmissions from node $(n_i,l_i-1)$ to node $(n_i,l_i)$, and $\lambda _{(n_i,l_i-1)}^{(n_i,l_i)}$ is the transmission delay between these two nodes. Here, the expectation $\mathbb{E}_s[\cdot]$ is taken over all the possible system states.
 
\begin{figure}[h!]
    \centering
    \includegraphics[width=0.5\textwidth]{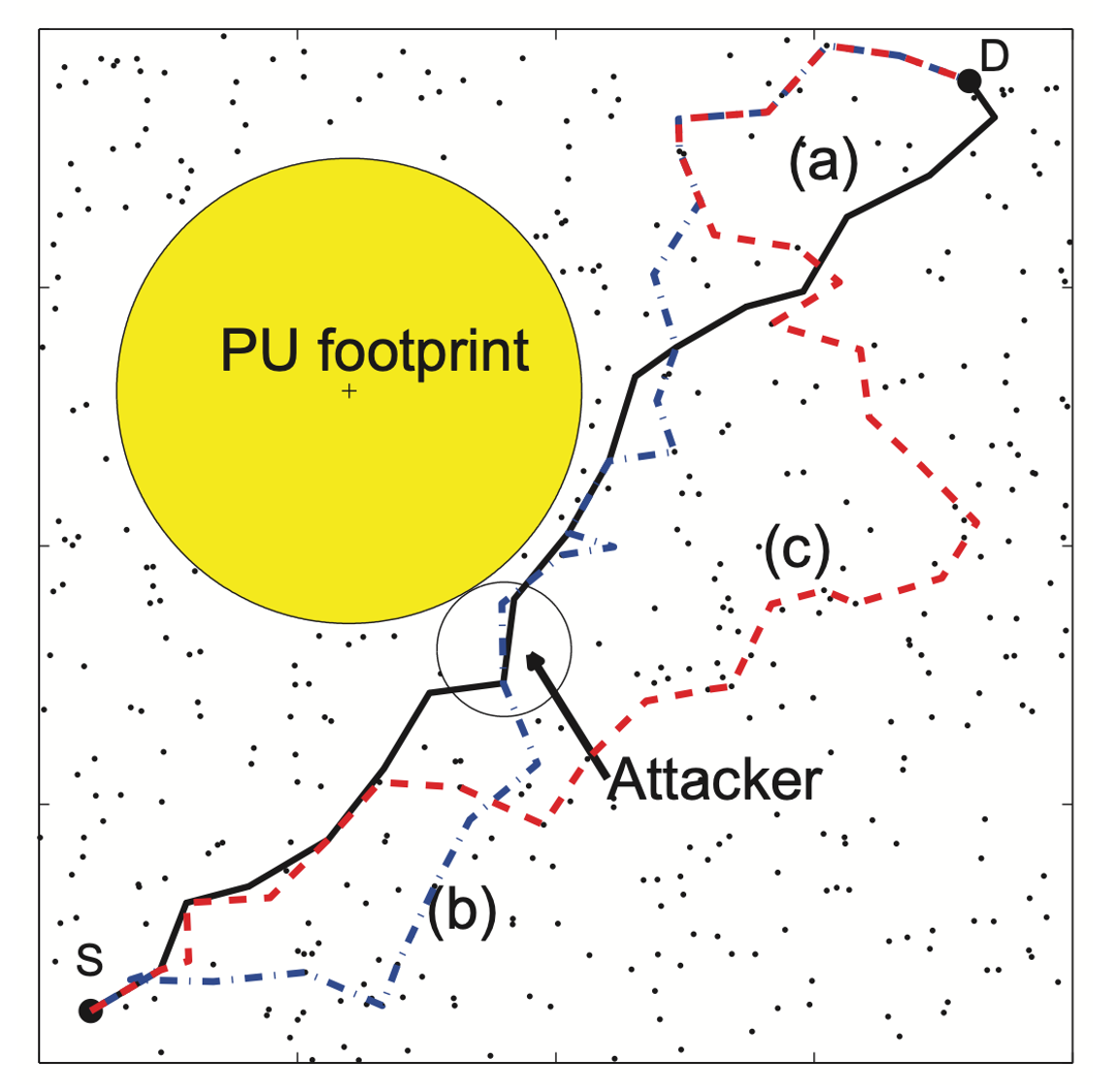}
    \caption{Illustration of a random network topology for 500 secondary users with a source (S) and a destination (D), and routes of AODV and the proposed secure routing algorithm in 2 km by 2 km area. The PU footprint denotes the set of nodes unavailable to secondary users. Without an attacker, AODV establishes the route path (a), described by the solid line, while the route path (b), the blue dashed line, is generated by the Boltzmann-Gibbs learning method. Even though the AODV path is the shortest path between the source and the destination, it is disrupted by malicious attacks. By contrast, the learning method can develop a new route path (c) that circumvents jammers, leading to a resilient routing mechanism.    }
    \label{fig:routing}
\end{figure}

Due to a lack of complete knowledge of adversaries and payoff structures,  Boltzmann-Gibbs reinforcement learning \eqref{eq:sbr} is utilized to find the optimal path because of its capability of estimating the expected utility.  The resulting secure routing algorithm can spatially circumvent jammers along the routing path and learn to defend against malicious attackers as the
state changes. As shown in \Cref{fig:routing}, the routing path generated from the proposed routing algorithm in \cite{zhu2011dynamic,song2019performance} can avoid the nodes compromised by the jammers. Thus, the routing algorithm stemming from the proposed game-theoretic formulation provides more resilience, security, and agility than the ad-hoc on-demand distance vector (AODV) algorithm, as AODV fails to dynamically adjust the routing path in the case of a malicious attack. Moreover, the proposed routing algorithm can reduce the delay time incurred by the attack due to its adaptive and dynamic feature, and hence, is more efficient than AODV.

\subsection{The Smart Grid}
\begin{figure}
    \centering
    \includegraphics[width=\textwidth]{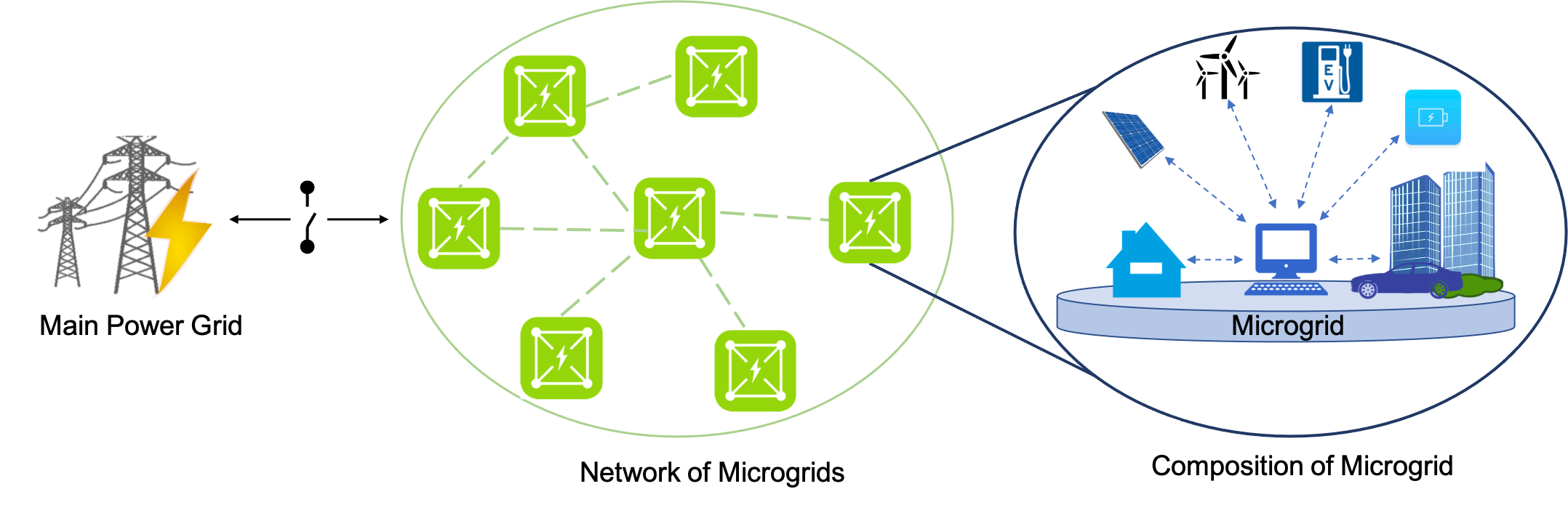}
    \caption{The integration of microgrids. A microgrid consists of a controller, consumers, generators, and energy storage. In the grid, microgrids can either be connected to the main grid or other microgrids, and these networked microgrids can operate, communicate, and interact autonomously to deliver power and electricity to their consumers efficiently. }
    \label{fig:smart_grid}
\end{figure}

Gradual replacements of conventional energies with renewable energies greatly help with the reduction of greenhouse gases and the mitigation of climate change. More and more microgrids are being integrated with the main power grid, which are green systems that rely on renewable distributed resources such as wind turbines and fuel cells. As shown in \Cref{fig:smart_grid}, the integration of microgrids can enhance the stability, resiliency, and reliability of the power system, as they can operate independently from the main power grid autonomously. Such integration, together with smart meters and appliances, produces the so-called smart grid, a modern infrastructure for the reliable delivery of electricity.    

The future smart grid is envisioned as a large-scale cyber-physical system comprising advanced power, communications, control, and computing technologies. To accommodate these technologies employed by different parties in the grid and to ensure an efficient and robust operation of such heterogeneous and large-scale cyber-physical systems, game-theoretic methods have been widely employed in smart grid management problems.  In the grid, microgrids are modeled as self-interested players who can operate, communicate, and interact autonomously to deliver power and electricity to their consumers efficiently. Here, we discuss a microgrid management mechanism developed in \cite{chen2016game}, built on game-theoretic learning, enabling autonomous management of renewable resources.

\begin{figure}[]
    \centering
    \includegraphics[width=\textwidth]{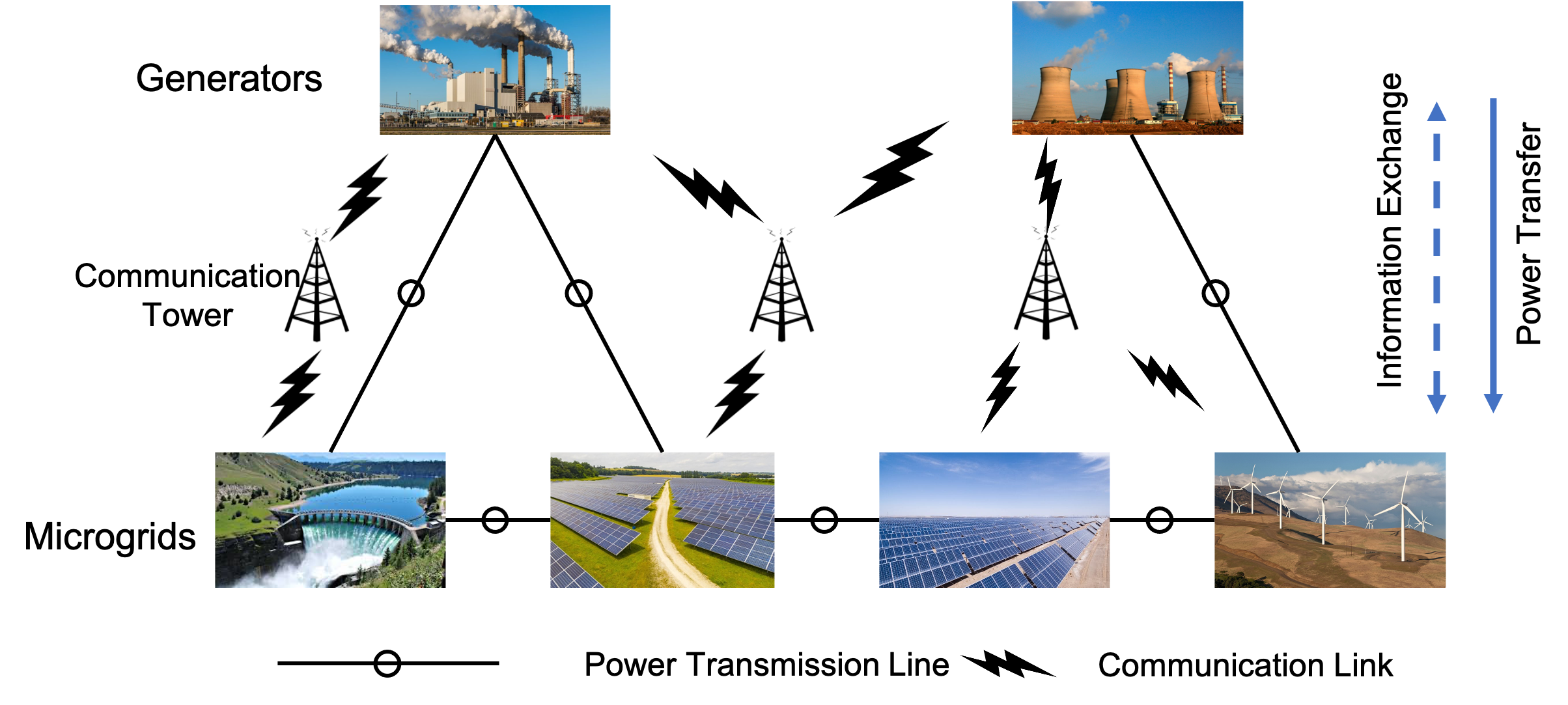}
    \caption{Smart grid hierarchy model. The upper layer containing conventional generators forms a generator network, and the distributed renewable energy generators in the bottom layer constitute the microgrid network; the information exchange, such as the electricity market price and the amount of power generation, between the two layers are through the communication network layer in the middle. }
    \label{fig:powerplant}
\end{figure}

The system model considered in \cite{chen2016game} includes the generators, microgrids, and communications. As shown in \Cref{fig:powerplant}, generators in the upper layer determine the amount of power to be generated, along with the electricity price, and send them to the bottom layer. A microgrid can generate renewable energies and make decisions by responding to the strategies of the generators and other microgrids to optimize their payoffs, specified in the following game-theoretic model.

Let $\mathcal{N}_d=\{r,1,2,...,N_d\}$ be the set of $N_d+1$ buses in a power grid, where $r$ denotes the slack bus. Assume that a smart grid is composed of load buses and generator buses and let $p^{\rm g}_i$, $p^{\rm l}_i$ and $\theta_i$ be, respectively, the power generation, power load, and voltage angle at the $i$-th bus. Note that the active power injection at the $i$-th bus satisfies 
\begin{equation*}%\label{activepower}
    p_i=p_i^{\rm g}-p_i^{\rm l},\quad \forall\ i\in\mathcal{N}_d,
\end{equation*}
while the balance of the grid gives $\sum_{i\in\mathcal{N}_d}p_i^{\rm g}=\sum_{i\in\mathcal{N}_d}p_i^{\rm l}.$ Let $\mathcal{N} := \{1, 2,...,N\} \subseteq \mathcal{N}_d$ be the set of $N$ buses that can generate renewable energies, such as wind power, solar power, etc. 
%Denote the set of rest buses by  $\underline{\mathcal{N}}:=\{N +1,N +2,...,\Bar{N}\}=\Bar{\mathcal{N}}\backslash\mathcal{N}$. 

In the game considered in \cite{chen2016game}, the utility function of the $i$-th bus measures not only economic factors related to power generation but also the efficiency of the microgrids. Before giving the mathematical definition of the utility function, we first introduce the following notations. Let $c_i$ be the unit cost of generated power for the $i$-th player, and $c$ the unit price of renewable energy for sale defined by the power market. $c_i,c$ are quantities relevant to the profit gained by the bus. For the efficiency part, denote by $r_i$ a weighting parameter that measures the importance of regulations of voltage angle at the $i$-th bus. Further, $[s_{ij}]_{i,j\in\mathcal{N}_d}=-[b_{ij}]_{i,j\in\mathcal{N}_d}^{-1}$, where $b_{ij}$ is the imaginary part of the element $(i, j)$ in the admittance matrix of the power grid. Moreover, each microgrid has a maximum generation, denoted by $\Bar{p}_i^{\rm g}$. Finally, we note that as a physical constraint, $[s_{ij}]$ and $[p_i]$ satisfy \eqref{eq:power_flow}  due to the power flow equation \cite{chen2016game}
\begin{align}\label{eq:power_flow}
    \sum_{j \in \mathcal{N}_d\backslash\mathcal{N} } s_{i j} p_{j}+\sum_{j \neq i \in \mathcal{N}} s_{i j} p_{j}=\theta_{i}-s_{i i} p_{i}, \forall i \in \mathcal{N},
\end{align}
where $\theta_i$ is the voltage angle of the $i$-th bus. With all the notations above, the utility function of the $i$-th bus is defined as 
\begin{equation*}
\begin{aligned}
    {u}_i(p_{i}^{\rm g},p_{-i}^{\rm g}):=-c_ip_{i}^{\rm g}-c\left(p_{i}^{\rm l}-p_{i}^{\rm g}\right)-\frac{1}{2}r_i^2\left(\sum_{j\in\mathcal{N}_d}s_{ij}p_j\right),\quad 0\leq p_i^{\rm g}\leq \Bar{p}_i^{\rm g},\quad i\in\mathcal{N}.
\end{aligned}
\end{equation*}

Three learning methods are proposed in the paper to seek the Nash equilibrium, all based on best response dynamics \eqref{eq:br_dis}. The first two algorithms are parallel-update algorithm (PUA) and random-update algorithm (RUA) studied in \cite{alpcan2002cdma}. PUA is essentially the best response algorithm we represent in \eqref{eq:br_dis}, with the learning rate $\lambda_i^k$ being zero for all $i$, and all players update their strategies in parallel.  As its name suggests, RUA incorporates randomness into the best response algorithm, resulting in an $\epsilon$-greedy best response algorithm: players update their strategies according to \eqref{eq:br_dis} with probability $1-\epsilon$, with $ \epsilon\in(0,1)$ and retain their previous strategies otherwise. When $\epsilon=0$, players constantly update their strategies in every round; in this case, RUA reduces to PUA. 

\begin{figure}[!ht]
    \includegraphics[width=\textwidth]{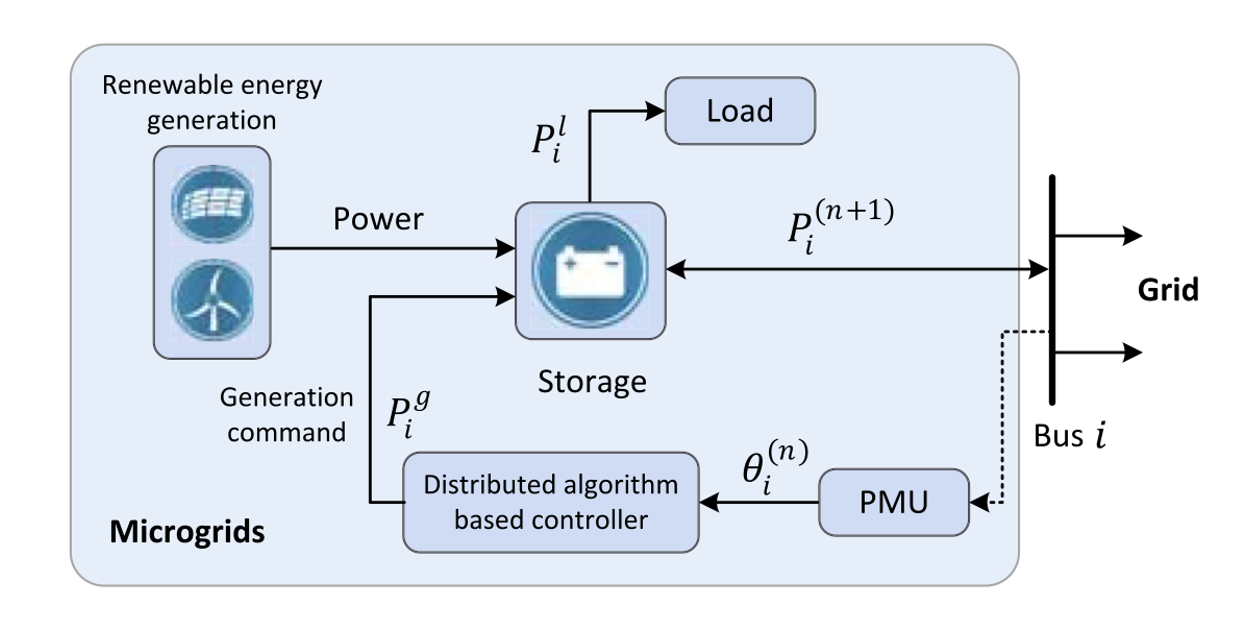}
    \caption{The framework to implement the PMU-enabled distributed algorithm.
PMU measures the voltage angle at the bus, and the controller generates a
command regarding the amount of microgrid renewable energy injection from the local storage to the grid based on the received voltage angle.}
    \label{fig:PDA}
\end{figure}

However, as special cases of \eqref{eq:br_dis},  PUA and RUA require global information regarding the grid, including the specific generated power of generators and other players' active power injections, which are assumed to be private in practice. Hence, to implement these algorithms, communication networks are needed to broadcast information to players, which is costly and not confidential. As a possible remedy, we can consider incorporating utility estimation and using smoothed best response dynamics \eqref{eq:sbr} as in the wireless setting. Another more straightforward approach, as shown in the paper, is to modify the best response algorithm by using the power flow equations in the smart grid.  Based on a phasor measurement unit (PMU), the third algorithm, termed PMU-enabled distributed algorithm (PDA), enables each player to compute the aggregation of others' actions, and the only information needed is the player's voltage angle $\theta_i$. Therefore, by taking into account the power flow equation \eqref{eq:power_flow}, a player does not need other players' private information of active power injection when using PDA, as shown in \Cref{fig:PDA}. Compared with the other two, PDA requires much less information and is more self-dependent as players only need their current voltage angles $\theta_i$, and the common knowledge of the electricity price.  

% By acquiring the knowledge of the current voltage angle $\theta_i^{(k)}$ and using the power flow equation
% \begin{equation*}
%     \sum_{j\in\underline{N}}s_{ij}p_j-\sum_{j\in{N}\backslash\{i\}}s_{ij}p_j=\theta_i-s_{ii}p_i,\quad \forall\ i\in\mathcal{N},
% \end{equation*}
% the players can simplify the best response update as the following
% \begin{equation*}
%     p^{(k+1)}_i=\min\left\{p^{\max}_i,\max\left\{-p_i^{\rm l},\frac{1}{s_{ii}}\left(a_i-\theta_i^{(k)}+s_{ii}p_i^{(k)}\right)\right\}\right\},
% \end{equation*}
% which is also guaranteed to converge to the unique Nash equilibrium under the same condition as PUA.

As indicated in \cite{chen2016game}, the effectiveness and resiliency of the algorithm have been validated via case studies based on the IEEE 14-bus system: the game-theory-based distributed algorithm not only can converge to the unique Nash equilibrium but also provides strong resilience against fault models (generator breakdown, microgrid turn-off, and open-circuit of the transmission line, etc.) and attack models (data injection attacks, unavailability of PMU data and jamming attacks, etc.). The strong resilience enables the microgrids to operate appropriately in unanticipated situations. Moreover, the distributed algorithm enables autonomous management of renewable resources and the plug-and-play feature of the smart grid. The proposed learning algorithm only requires the players to have common knowledge without revealing their private information, which increases security and privacy and reduces communication overhead.

\subsection{Distributed Machine Learning over Networks}
The rise of Big Data has led to new demands for large-scale machine learning systems that promise adequate capacity to digest massive data sets and offer powerful predictive analytics. With the unrestrainable growth of data, large-scale machine learning needs to address new challenges regarding the scalability and efficiency of learning algorithms concerning computational and memory resources. Compared with classical machine learning approaches that are designed to learn from a single integrated data set, one of the promising research lines of large-scale machine learning is distributed machine learning over networks (DMLON), which aims to develop efficient and scalable algorithms with reasonable requirements of memory computation resources, by allocating the learning processes among several networked computing units with distributed data sets. 

The key feature of DMLON is that data sets are stored and processed locally on these computing units, which enables distributed and parallel computing schemes in large-scale machine learning systems. Compared with centralized approaches, distributed machine learning avoids maintaining and mining a central data set and preserves data privacy, as these networked units exchange knowledge about the learned models without exchanging raw private data. 

Based on the idea of ``local learning and global integration,'' DMLON can utilize different learning processes to train several models from distributed data sets and then produce an integration of learning models that can increase the possibility of achieving higher accuracy, especially on a large-size domain. For example, in federated learning \cite{konevcny2016federated}, the global integration is created by a third-party coordinator other than computing units, which makes networked computing units collaboratively train a machine learning model using their data in security. On the other hand, as indicated in \cite{liu2020communication}, such a global integration can also stem from the collective patterns of local learning without external enforcement. The key behind this bottom-up integration is that each computing unit is modeled as a self-interested player who learns the learning model based on the local data set and the feedback from its neighbors. It has been shown in the paper that by modeling DMLON as a noncooperative game, game-theoretic learning methods lead to a communication-efficient distributed machine learning, where the global outcome is characterized by the Nash equilibrium, resulting from players' self-adaptive behaviors.  

Specifically, the networked system of computing units is described by a graph with the set of nodes $\mathcal{N}_m:=\{1,2,\ldots, N\}$ representing these units. Each node $i\in \mathcal{N}_m$ possesses local data that cannot be transferred to other nodes.  In the game model considered in \cite{liu2020communication, liu2023}, instead of fixing the network topology, nodes can determine the network's connectivity based on their attributes when they perform learning tasks, resulting in a network formation game. In mathematical terms, the action of node $i$ consists of two components: the learning parameter $\theta_i\in \R^d$, and the network formation parameter $e_i\in \R^{N-1}$. The first component $\theta_i$ corresponds to the weights or parameters of the machine learning model, which captures the local learning process at node $i$, and the corresponding empirical loss, given the local data, is denoted by $L_i(\theta_i)$. In addition to this learning parameter $\theta_i$, the network formation parameter $e_i$ plays an important role in bringing up the global integration. The parameter $e_i:=(e_i^j)_{j\neq i, j\in \mathcal{N}}\in [0,1]^{N-1}$ denotes concatenation of weights on the directed edges from node $i$ to other nodes, where $e_i^j$ can be interpreted as the attention node $i$ pays to the local learning at node $j$, and this further influence the communication between the nodes. Each node can communicate with its neighbors during the distributed learning process to exchange learning parameters if their objectives are aligned. Otherwise, the corresponding edge weight $e_i^j$ is set to zero. For node $i$, the communication cost is  $C_i(\theta_i,\theta_{-i},e_i)$.
In the game considered in \cite{liu2020communication}, each node aims to maximize its utility function, defined as 
\begin{align*}
    u_i(\theta_i,\theta_{-i},e_i,e_{-i}):=-L_i(\theta_i)-C_i(\theta_i,\theta_{-i},e_i),
\end{align*}
In this definition, the first term $L_i(\theta_i)$ captures the local learning process at node $i$, whereas the second term $C_i(\theta_i,\theta_{-i},e_i)$  depicts the interactions among nodes. The objective of each node is to improve the performance of learning while reducing the communication overhead.  

\begin{figure}
    \centering
    \includegraphics[width=0.8\textwidth]{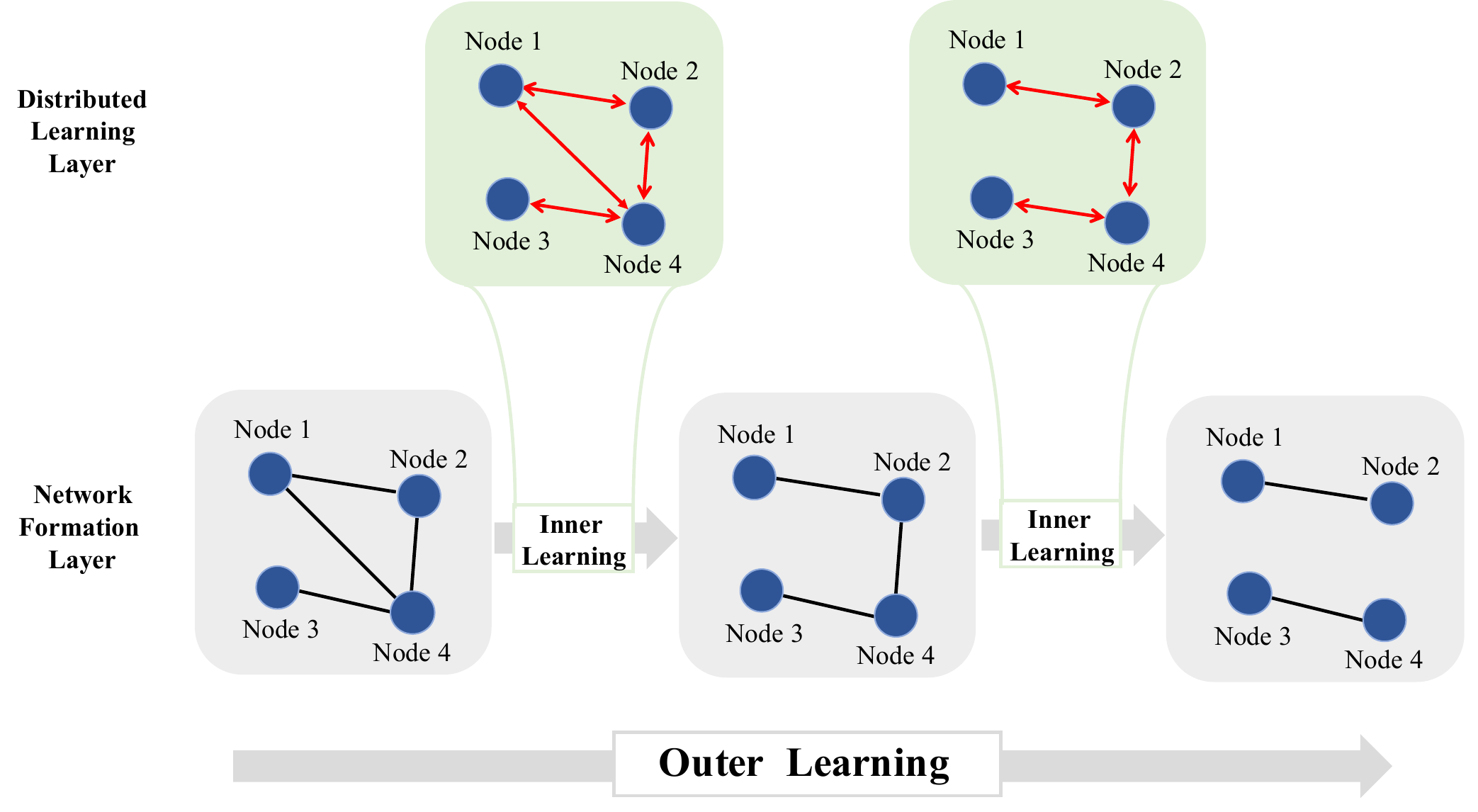}
    \caption{A schematic representation of two-layer learning. The directed red lines stand for the communication between nodes. In the network formation layer, the nodes learn to eliminate/establish links with other nodes to achieve efficient communication. In the distributed machine learning layer, the nodes communicate their parameters with their neighbors and perform their learning tasks.}
    \label{fig:process}
\end{figure}

A two-layer learning approach is proposed in \cite{liu2020communication} to find the Nash equilibrium of the game, and a schematic representation is provided in \Cref{fig:process}. The outer layer corresponds to network formation learning, where each node decides its network formation parameter $e_i$ with the learning parameter fixed, and the joint parameters of all nodes $e=(e_i)_{i\in \mathcal{N}_m}$ give rise to new network topology, leading to efficient communication. In network formation learning, each node decides their optimal parameter $e_i$ by gradient play \eqref{eq:pgd}, and computing the individual payoff gradient  $\nabla_{e_i} u_i(\theta_i,\theta_{-i},e_i,e_{-i})$ relies on the stabilized learning parameters $\theta_i,\theta_{-i}$ given by the inner layer: distributed learning layer.  In this inner learning, the network formation parameter is fixed, and each node implements online mirror descent \eqref{eq:omd} for seeking the Nash equilibrium with the local feedback under the current network topology, as the networked nodes can exchange information with their neighbors.  

Compared with existing works on distributed machine learning, the game-theoretic method studied in \cite{liu2020communication} enables distributed machine learning over strategic networks. On the one hand, the global outcome characterized by the Nash equilibrium is self-enforcing, resulting from the coordinated behaviors of independent computing compared with the external enforcing one in federated learning. This bottom-up approach scales efficiently when additional computing units are introduced into the system. On the other hand, the strategic interactions over the network, described by the network formation decision of each node, create a network intelligence that allows each computing unit to adaptively adjust the underlying topology, resulting in a desired distributed learning pattern that minimizes communication costs during the learning process.  

\subsection{Emerging Network Applications}
From the examples above, game-theoretic learning provides a natural scalable design framework to create network intelligence for autonomous control, management, and coordination of large-scale complex network systems with heterogeneous parties. In the following, we offer some thoughts regarding various applications of game-theoretic learning in a broader context, showing that such a design framework is pervasive for diverse network problems.

Interdependent infrastructure networks, including wireless communication networks and the smart grid, play a significant role in modern society, where Internet-of-Things (IoT) devices are massively deployed and interconnected. These devices are connected to cellular/cloud networks, creating multi-layer networks, referred to as networks-of-networks \cite{juntao19infra}. The smart grid is one prominent example, where wireless sensors collect the data of buses and power transmission lines, forming a sensor network built on the power networks for grid monitoring and decision planning purposes \cite{zhu2019multilayer}. Besides, the networks-of-networks model has also been extensively studied in other infrastructure networks. For instance, in an intelligent transportation network, apart from vehicle-to-vehicle (V2V) communications, vehicles can also communicate with roadside infrastructures or units belonging to one or several service providers to exchange various types of data related to different applications, such as GPS navigation. In this case, the vehicles form one network while the infrastructure nodes form another network, and the interconnections between the two networks lead to the intelligent management and operation of modern transportation networks.

Due to interdependent networks' heterogeneous and multi-tier features, the required management mechanisms or controls can vary for different networks. For example, the connectivity of sensor networks in smart grids or V2V communication networks requires higher security levels than the infrastructure networks, as cyberspace is more likely to be targeted by adversaries \cite{farooq18iobt}. Therefore, to manage and secure interdependent infrastructure networks, game-theoretic learning methods, especially heterogeneous learning \cite{zhu10heter,zhu12heter}, can be used to design decentralized and resilient mechanisms that are responsive to attacks and adaptive to the dynamic environment, as different parties in interdependent infrastructure networks may acquire different information.  For further readings on this topic, we refer the reader to \cite{zhu12heter,juntao19infra} and references therein.

Similar to distributed optimization and machine learning based on game-theoretic learning, the control of autonomous mobile robots can also be cast as a Nash equilibrium seeking problem over networks, where the equilibrium is viewed as the desired coordination of all robots \cite{kaminka10,inujima13}. For applications of this kind, where the nature of robot movements determines the network topologies, dynamic games over networks are considered, and corresponding learning algorithms are employed. Based on their observations of the surroundings, robots rely on game-theoretic learning, for example, reinforcement learning, for developing self-rule policies, leading to a need for decentralized and scalable control laws for multi-agent robotic systems. Moreover, reinforcement learning has proven effective for real-world multi-agent robotic control when combined with powerful function approximators, such as deep neural networks. This area of research, termed deep multi-agent reinforcement learning \cite{kaiqing_overview,foerster2016learning}, is growing rapidly and attracting the attention of researchers from machine learning, robotics as well as control communities.

In addition to these prescriptive mechanisms in engineering practices, game-theoretic learning also provides a descriptive model for studying human decision-making and strategic interactions in epidemiology and social sciences, where the Nash equilibrium represents a stable state of the underlying noncooperative game. For example, a differential game model has been proposed in \cite{huang2019differential} to study the virus or diseases spreading over the network. Authors have developed a decentralized mitigation mechanism for controlling the spreading. Such an approach has been further explored in \cite{chen2020optimal}, where an optimal quarantining strategy of suppressing two interdependent epidemics spreading over complex networks has been proposed and proven robust against random changes in network connections.

\section{Summary}\label{sec:concl}
This article provides a comprehensive overview of game theory basics and related learning theories, which serve as building blocks for systematically treating multi-agent decision-making over networks.  We have elaborated on the game-theoretic learning methods for network applications drawn from spanning emerging areas such as the next-generation wireless networks, the smart grid, and networked machine learning. In each area, we have identified the main technical challenges and discussed how game theory can be applied to address them in a bottom-up approach.

From the surveyed works, we conclude that noncooperative game theory is the cornerstone of decentralized mechanisms for large-scale complex networks with heterogeneous entities, where each node is modeled as an independent decision-maker. The resulting collective behaviors of these rational decision-makers over the network can be mathematically depicted by the solution concept: Nash equilibrium. In addition to various game models, learning in games is of great significance for creating distributed network intelligence, which enables each entity in the network to respond to unanticipated situations, such as malicious attacks from adversaries in cyber-physical systems \cite{zhu2019multilayer}. Under local or individual feedback, the introduced learning dynamics lead to a decentralized and self-adaptive procedure, resulting in desired collective behavior patterns without external enforcement.

Beyond the existing successes of game-theoretic learning, which mainly focuses on learning in static repeated games,  it is also of interest to investigate dynamic game models and associated learning dynamics, in order to better understand the decision-making process in dynamic environments. The motivation for studying dynamic models and related learning theory stems, on the one hand, from the pervasive presence of time-varying network structures, such as generation and demand in the smart grid \cite{chen2016game}.  On the other hand, by defining auxiliary state variables, the problem of decision-making under uncertainties can be modeled as a dynamic game, where the state of the game includes the hidden information players do not have access to when making decisions. For example, the state variable can capture the uncertainty of the environment, as we have discussed in the context of the dynamic routing problem \cite{zhu2011dynamic}, or it can describe the global status of the entire system, as we have shown in the example of distributed optimization \cite{li2013designing}. The dynamic game models not only simplify the construction of players' utilities and actions, providing a clear picture of the strategic interactions under uncertainties in the dynamic environment but can also offer a scalable design framework for prescribing players' self-adaptive behaviors that lead to equilibrium states under various feedback structures.        

To recap, this article has presented a comprehensive overview of game-theoretic learning and its potential for tackling the challenges emerging from network applications.  The combination of game-theoretic modeling and related learning theories constitutes a powerful tool for designing future data-driven network systems with distributed intelligent entities, which serve as the bedrock and a key enabler for resilient and agile control of large-scale artificial intelligence systems in the near future. 
\newpage
\bibliographystyle{ieeetr}
\bibliography{reference}
\begin{appendix}

\section{ Fictitious Play }\label{side:fp}
Consider the repeated play between two players, with each player knowing his own utility function. Further, each player is able to observe the actions of the other player and choose an optimal action based on the empirical frequency of these actions.  

In fictitious play, from player 1's viewpoint, player 2's strategy at time $k$ can be estimated as $\pi_2^k(a)=\sum_{s=1}^k\mathbbm{1}_{\{a_2^s=a\}}/k, a\in\mathcal{A}_2$, which is the empirical frequency of actions player 2 has implemented up to that point. $\pi_2^k$ can be computed by a moving average scheme: $${\pi}_2^{k}=(1-\frac{1}{k}){\pi}_2^{k-1}+\frac{1}{k}e_{a_2^k}.$$ Using this, player 1 chooses the best response:  $a_1^{k+1}= \argmax_{a\in \mathcal{A}_1}u_1(a,\pi^k_2)$ for the next play. Then, the empirical frequency of player 1's implemented actions is updated according to 
$$\pi_1^{k+1}=(1-\frac{1}{k+1})\pi_1^{k}+\frac{1}{k+1}e_{a_1^{k+1}},$$ where $e_{a_1^{k+1}}\in \Delta(\mathcal{A}_1)$ is exactly given by $BR_1(\pi_2^{k})$ and the equation is the same as the one in \eqref{eq:br_dis}, with the learning rate being $\lambda_1^{k}=\frac{1}{k+1}$. Hence, we conclude that in fictitious play, a player's empirical play follows best response dynamics. Furthermore, if we replace the best response mapping $BR$ with the quantal response $QR^\epsilon$, we then obtain  an important variant: stochastic fictitious play \cite{fuden_learning}.

\section{ Replicator Dynamics }\label{side:rd}
Recall that continuous-time learning dynamics under dual averaging is  
\begin{align}
\begin{aligned}
	&\frac{d {\hat{\mathbf{u}}}_{i}(t)}{dt}=\mathbf{u}_i(\pi_{-i}(t)),\\
	& \pi_i(t)=QR^\epsilon(\hat{{\mathbf{u}}}_i(t)).
\end{aligned} \tag{\text{DA-c}}
\end{align}
We now consider the entropy regularizer $h(x)=\sum_{x_i}x_i\log x_i$ and let $\epsilon=1$ for simplicity. Differentiate the strategy $\pi_{i}(t)$ with respect to time variable in \eqref{eq:da_con}, arriving at  
\begin{align}\label{eq:rd}
	\frac{d\pi_{i,a}(t)}{dt}&=\frac{1}{(\sum_{a'}e^{\hat{\mathbf{u}}_{i,a}(t)})^2}\left(\frac{d\hat{\mathbf{u}}_{i,a}(t)}{dt} e^{\hat{\mathbf{u}}_{i,a}(t)}\sum_{a'}e^{\hat{\mathbf{u}}_{i,a'}(t)}-e^{\hat{\mathbf{u}}_{i,a}(t)}\sum_{a'}e^{\hat{\mathbf{u}}_{i,a'}(t)}\frac{d\hat{\mathbf{u}}_{i,a'}(t)}{dt}\right)\nonumber\\
	&=\pi_{i,a}(t)\left(\frac{d\hat{\mathbf{u}}_{i,a}(t)}{dt}-\sum_{a'}\pi_{i,a'}(t)\frac{d\hat{\mathbf{u}}_{i,a'}(t)}{dt}\right)\nonumber\\
	&=\pi_{i,a}(t)[u_i(a,\pi_{-i}(t))-u_i(\pi_i(t),\pi_{-i}(t))].\tag{\text{RD}}
\end{align}

From the equation above, we can see that for a certain action $a$, if its outcome $u_i(a,\pi_{-i}(t))$ is above the average $u_i(\pi_i(t),\pi_{-i}(t))$, then it will be ``reinforced'' in the sense that the probability of choosing $a$ gets higher as time evolves. The above equation \eqref{eq:rd} is referred to as  replicator dynamics, and has been widely used in evolutionary game theory to understand natural selection and population biology. We consider a two-population system and we reinterpret the elements in the two-player game using population biology language. For population 1, there are $|\mathcal{A}_1|$ types and each type is specified by an element $a\in \mathcal{A}_1$.  We let $\pi_{1,a}(t)$ be the percentage of type $a$ in population 1 at time $t$, and assume here that $\pi_1(t)$ is differentiable with respect to time $t$, as the population, which is infinitely large, interacts with the other population in a continuous-time manner. 
	
For population 2, we have similar notions. If individuals from the two population meet randomly, then they engage in a competition or a game with payoff dependent on their types. For example, if type $a_1$ from population 1 competes with type $a_2$ from population 2, then the payoffs for the two types are given by  $u_1(a_1,a_2)$ and $u_2(a_1,a_2)$, respectively. For population $i$, if we assume that the per capita rate of growth is given by the difference between the payoff for type $a$ and the average payoff in the population, a rule studied in \cite{maynard73logic}, 
then the percentage of different types within a population is precisely described by 
$$\frac{1}{\pi_{i,a}}\frac{d\pi_{i,a}(t)}{dt}=u_i(a,\pi_i(t))-u_i(\pi_i(t),\pi_{-i}(t)),$$  which is exactly the replicator dynamics \eqref{eq:rd}. In addition, as shown in \cite{PM16lyapnov}, different regularizers lead to different learning dynamics, which display different asymptotic behavior accounts for the evolutionary process under different circumstances. 

With replicator dynamics and other related evolutionary dynamics, biologists can predict the evolutionary outcome of the multi-population system by examining the Nash equilibrium of the underlying game, which brings strategic reasoning into population biology and has a profound influence in evolutionary game theory \cite{cressman14replicator,sandholm10evolu_game}. 
Moreover, the Nash equilibrium in this population game, characterized by the limiting behavior of the dynamics under proper conditions \cite{cressman14replicator}, represents an evolutionarily stable state of the population, which is an important refinement of Nash equilibrium. When this stable state is reached, natural selection alone is sufficient to prevent the population from being influenced by mutation \cite{sandholm10evolu_game,hofbauer03evolu_dyna}.  For more details on this refinement and its application in biology, we refer the reader to \cite{basar,cressman14replicator,sandholm10evolu_game,hofbauer03evolu_dyna}.

\section{ Stochastic Approximation Theory}
\setcounter{equation}{0}
\renewcommand{\theequation}{C\arabic{equation}}
\label{side:sa}
Following the multiple timescale stochastic approximation framework developed in \cite{Borkar:2009ts,yin03stochastic_aaprox}, one can write \eqref{eq:utility} and \eqref{eq:strategy}  using discrete-time stochastic approximation
\begin{align}
\begin{aligned}
	&\pi_i^{k+1}-\pi_i^{k}= \bar{\lambda}_i^k\left(f_i(\pi_i^k,\hat{\mathbf{u}}_i^{k+1})+M_i^{k+1}\right),\\
	&\hat{\mathbf{u}}_i^{k+1}-\hat{\mathbf{u}}_i^{k}= \bar{\mu}_i^k \left(g_i(\pi_i^k,\hat{u}_i^k)+\Gamma_i^{k+1}\right),
\end{aligned}\label{eq:dsa_i}
\end{align}   
where $f_i(\pi_i^k,\hat{\mathbf{u}}_i^{k+1})$ and $g_i(\pi_i^k,\hat{\mathbf{u}}_i^k)$ are the mean-field components of \eqref{eq:utility} and \eqref{eq:strategy}, respectively, and are defined as 
\begin{align*}
    f_i(\pi_i^k,\hat{\mathbf{u}}_i^{k+1})&=\mathbb{E}[F_i(\pi_i^k,\hat{\mathbf{u}}_i^{k+1}, U_i^{k+1}, a_i^{k+1})|\mathcal{F}^{k-1}],\\
    g_i(\pi_i^k,\hat{\mathbf{u}}_i^k)&=\mathbb{E}[G_i(\pi_i^k,\hat{\mathbf{u}}_i^k, U_i^{k+1}, a_i^{k+1})|\mathcal{F}^{k-1}].
\end{align*}
With the mean-field part defined as above, $M_i^{k+1} = F_i(\pi_i^k,\hat{\mathbf{u}}_i^{k+1},U_i^{k+1}, a_i^{k+1})-f_i(\pi_i^k,\hat{\mathbf{u}}_i^{k+1})$ and $\Gamma_i^{k+1}$ takes a similar form. $\bar{\lambda}_i^k, \bar{\mu}_i^k$ are time-scaling factors dependent on the learning rates $\lambda_i^k, \mu_i^k$, which account for the adjustment of the original step sizes in asynchronous schemes \cite{Borkar:2009ts,Perkins12asy_SA}, and in synchronous cases, the time scaling factors coincide with the original step sizes. Similar to our discussion in the main text (see \eqref{eq:moving_dyna} and \eqref{eq:br_dyna}), we consider the dynamical system of the joint strategy profile $\pi^k$ and utility vector $\hat{\mathbf{u}}^k$
\begin{align}
\begin{aligned}
	&\pi^{k+1}-\pi^{k}= \bar{\lambda}^k\left(f(\pi^k,\hat{\mathbf{u}}^{k+1})+M^{k+1}\right),\\
	&\hat{\mathbf{u}}^{k+1}-\hat{\mathbf{u}}^{k}= \bar{\mu}^k \left(g(\pi^k,\hat{u}^k)+\Gamma^{k+1}\right),
\end{aligned}\tag{\mbox{\rm{DSA}}}\label{eq:dsa}
\end{align}   
where $f$ and $g$ are concatenations of $\{f_i\}_{i\in\mathcal{N}}$ and  $\{g_i\}_{i\in\mathcal{N}}$, respectively. $\bar{\lambda}^k, \bar{\mu}^k$ and $M^k, \Gamma^k$ take similar forms.

As we have discussed in ``\nameref{sec:conver}'', in order to obtain an approximately accurate score function, the two coupled dynamical systems in \eqref{eq:dsa} should operate on different timescales: the score function $\hat{\mathbf{u}}^{k}$ should be updated sufficiently many times until near-convergence before updating the strategy. This two-timescale iteration can be achieved by adjusting the time-scaling factors: $\bar{\lambda}^k$ and $\bar{\mu}^k$ are chosen so that $\lim_{k\rightarrow\infty}\bar{\lambda}^k/\bar{\mu}^k=0$. To understand this timescale system, it is instructive to consider a coupled continuous-time dynamical system, as suggested in \cite{Borkar:2009ts}
\begin{align}
\begin{aligned}
    &\frac{d \pi(t)}{d t}=  f(\pi(t), \hat{\mathbf{u}}(t)),\\
	&\frac{d \hat{\mathbf{u}}(t)}{dt} =  \frac{1}{\varepsilon}g(\pi(t), \hat{\mathbf{u}}(t)),
	\end{aligned}\label{eq:diff_varep}
\end{align}
where in the limit $\varepsilon$ tends to zero. Hence, $\hat{\mathbf{u}}(t)$ is fast transient while $\pi(t)$ is slow. Then, we can analyze the long-run behavior of the above coupled system as if the fast process is always fully calibrated to the current value of the slow process. This suggests investigating the ODE
\begin{align}\label{eq:slow_g}
    \frac{d \hat{\mathbf{u}}(t)}{dt} =  g(\pi, \hat{\mathbf{u}}(t)),
\end{align}
where $\pi$ is held fixed as a constant parameter.  Suppose \eqref{eq:slow_g} has a globally asymptotically stable equilibrium $\Lambda(\pi)$, where the mapping $\Lambda(\cdot)$ satisfies regularity conditions specified in \cite{benaim05SADI, Perkins12asy_SA}. Then, it is reasonable to expect $\hat{\mathbf{u}}(t)$ given by \eqref{eq:slow_g} to closely track $\Lambda(\pi)$. In turn, this suggests that the investigation into the coupled system \eqref{eq:diff_varep} is equivalent to the study of the single-timescale one
\begin{align}\label{eq:pi_equiva}
    \frac{d \pi(t)}{d t}=  f(\pi(t), \Lambda(\pi(t))),
\end{align}
which would capture the long-run behavior of $\pi(t)$ in \eqref{eq:diff_varep} to a good approximation \cite{Borkar:2009ts}.

 Informally speaking, to study the convergence of \eqref{eq:dsa}, we can relate its discrete-time trajectory to that of \eqref{eq:diff_varep}, which is further equivalent to $(\pi(t),\Lambda(\pi(t)))$ specified by  \eqref{eq:pi_equiva}. Therefore, we can apply Lyapunov stability theory to \eqref{eq:pi_equiva}, in order to derive the convergence results of the original discrete-time algorithm. We begin with the linear interpolation process of the discrete-time trajectory, which connects the discrete-time system \eqref{eq:dsa} and its continuous-time couterpart \eqref{eq:diff_varep},\eqref{eq:pi_equiva}. Under some regularity conditions \cite{benaim05SADI}, for $\{\pi^k\}$, the sequence generated by \eqref{eq:dsa}, we can construct the following continuous time process $\bar{\pi}(t):\R_{+}\rightarrow \Delta(\mathcal{A})$, based on the linear interpolation of $\{\pi^k\}$. Letting $\tau^0=0$ and $\tau^k=\sum_{s=1}^k\bar{\lambda}^{s}$, we define
\begin{align*}
	\bar{\pi}(t):=\pi^k+(t-\tau^k)\frac{\pi^{k+1}-\pi^k}{\tau^{k+1}-\pi^k}, \quad t\in [\tau^k, \tau^{k+1}).
\end{align*}    
Similarly, we can define a continuous-time process $\bar{\mathbf{u}}(t)$ corresponding to $\{\hat{\mathbf{u}}^{k}\}$. 

{\color{blue}}

As shown in \cite{benaim05SADI,Perkins12asy_SA}, such a linearly interpolated process $(\bar{\pi}(t), \bar{\mathbf{u}}(t))$ is closely related to the flow of the following differential equations:
\begin{align}
\begin{aligned}
	&\frac{d \pi(t)}{d t}=  f(\pi(t), \hat{\mathbf{u}}(t)),\\
	&\frac{d \hat{\mathbf{u}}(t)}{dt} =  g(\pi(t), \hat{\mathbf{u}}(t)).
\end{aligned}\label{eq:diff_incl}
\end{align}
 We note that \eqref{eq:diff_incl} is defined for ease of presentation, and the actual differential inclusion systems involves rearrangement of several terms; which we refer the reader to \cite{Perkins12asy_SA} for more details. Further, we denote the flow of \eqref{eq:diff_incl} by $$\Phi_t(\pi^0,\mathbf{u}^0):=\{(\pi(t), \hat{\mathbf{u}}(t))| (\pi(t), \hat{\mathbf{u}}(t)) \text{ is a solution to \eqref{eq:diff_incl}, with } \pi(0)=\pi^0, \hat{\mathbf{u}}(0)=\mathbf{u}^0\}.$$
The key of stochastic approximation theory lies in the fact that in the presence of a global attractor for \eqref{eq:diff_incl}, the continuous-time process $(\bar{\pi}(t),\bar{u}(t))$ asymptotically tracks the flow with arbitrary accuracy over windows of arbitrary length  \cite{benaim05SADI}, 
\begin{align*}
	\lim_{t\rightarrow\infty}\sup_{s\in [0,T]}\operatorname{dist}\{(\bar{\pi}(t+s),\bar{\mathbf{u}}(t+s)),\Phi_s(\bar{\pi}(t),\bar{\mathbf{u}}(t))\}=0,
\end{align*}
where $\operatorname{dist}\{\cdot,\cdot\}$ denotes a distance measure on $\Delta(\mathcal{A})\times\R^{\mathcal{A}}$. We refer to $(\bar{\pi}(t),\bar{\mathbf{u}}(t))$ as an asymptotic pseudo-trajectory (APT) of the dynamics \eqref{eq:diff_incl}. In other words, in order to study the convergence of \eqref{eq:dsa}, we can resort to the convergence analysis of \eqref{eq:diff_incl}, which can be addressed by Lyapunov stability theory as shown in \cite{benaim05SADI, Perkins12asy_SA}, where the key conclusion is that if there is a global attractor $A$ for \eqref{eq:pi_equiva}. Then the interpolated process $(\bar{\pi}(t),\bar{\mathbf{u}}(t))$ or simply $(\pi^k, \bar{\mathbf{u}}^k)$ converges almost surely to $(A,\Lambda(A))$.

\end{appendix}

\end{document}